\documentclass[aps,pra,twocolumn,groupedaddress,showpacs,amsmath,amssymb,byrevtex]{revtex4}
\usepackage{graphicx}    % needed for including graphics e.g. EPS, PS
\usepackage{epstopdf}

\begin{document}

\title{Hanbury Brown-Twiss correlations and multimode dynamics of quenched spinor Bose-Einstein condensates}
\author{A. Vinit and C. Raman}
\email {chandra.raman@physics.gatech.edu}
\affiliation{School of Physics, Georgia Institute of Technology, Atlanta, Georgia 30332, USA}
\date{\today}

\begin{abstract}
We have studied the interaction between multiple, competing spatial modes that are excited by a quantum quench of an antiferromagnetic spinor Bose-Einstein condensate.  We observed Hanbury Brown-Twiss correlations and associated super-Poissonian noise in the mode populations.  The decay of these correlations was consistent with experimentally observed spin domain patterns.  Data were compared with a real-space Bogoliubov theory as well as numerical solution of the coupled Gross-Pitaevskii equations that were seeded by quantum noise via the truncated Wigner approximation.  The spatial modes that were both observed  experimentally and deduced theoretically are intimately connected to the inhomogeneous density profile of the condensate, which imparts many rich features to the dynamics.
\end{abstract}

\pacs{03.75.Mn,67.85.De,67.85.Fg,67.85.Hj}
\maketitle

\section{Introduction}

\begin{figure} [htbp]
\includegraphics[width= \columnwidth]{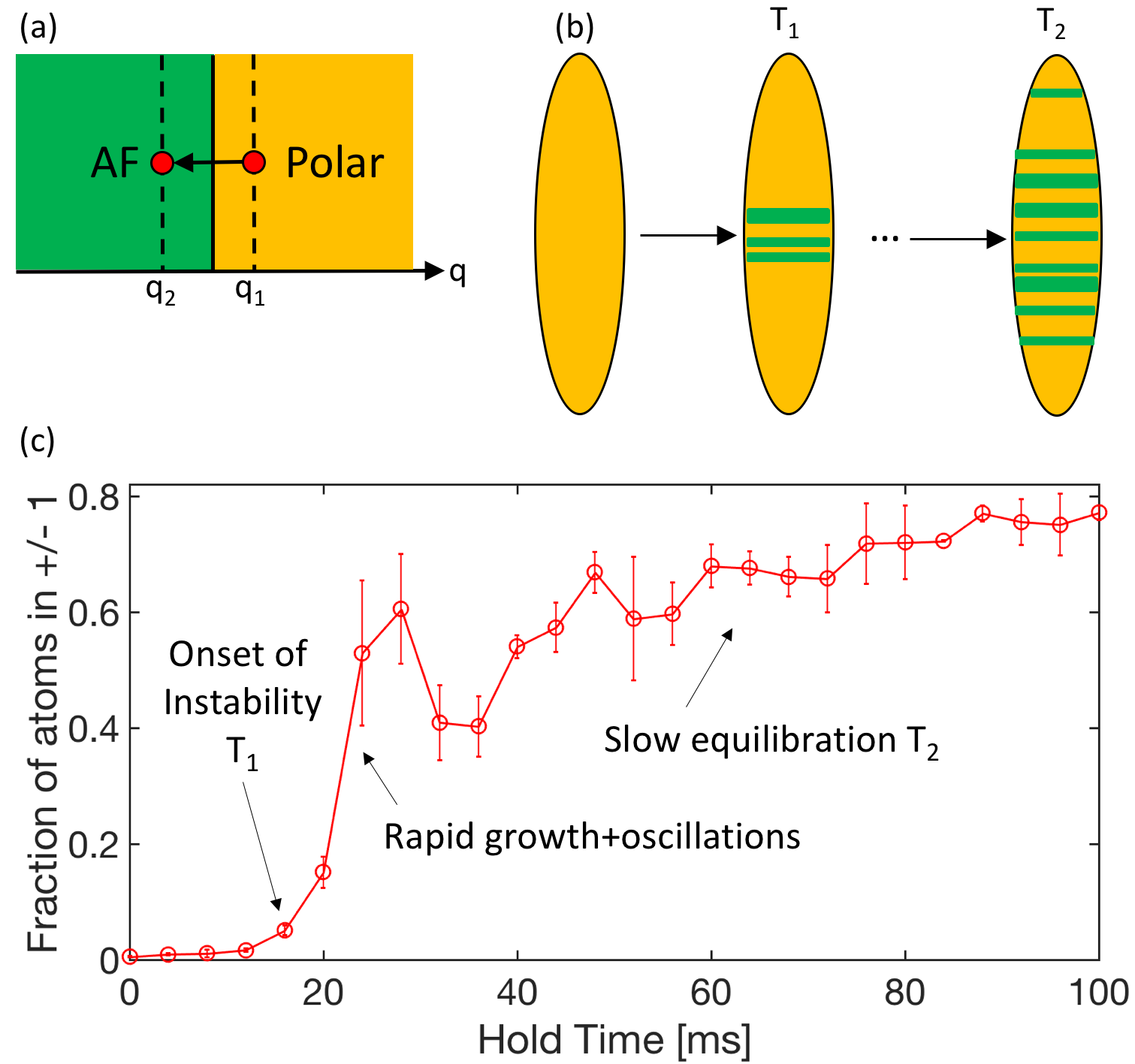}
\caption{(Color Online). Quantum quenches reveal two phases of an antiferromagnetic spinor BEC.  a) Quantum phase transition between polar and antiferromagnetic (AF) phases at zero quadratic Zeeman shift $q$. (b) Dynamical instability induced by a rapid quench from $q_1>0$ to $q_2<0$ results in the appearance of a small number of AF domains at short times $T_1$ after the quench, and a quasi-equilibrium between the phases at longer times $T_2$.  (c) Measurement of total fraction in the AF phase (the $m_F = \pm 1$ population fraction) versus time showing the various dynamical phases (figure adapted from \protect{\cite{vinit2013}}).  Error bars are the standard deviation of 3 separate measurements.}
\label{fig:diagram}
\end{figure}

Quantum phase transitions have become an important arena of exploration using ultracold atoms.  In contrast to the traditional view of these transitions that has focused on equilibrium properties \cite{sachdev99}, quantum gases afford a dynamical view made  experimentally possible by rapidly quenching the system \cite{langen2015}.  A major advantage exists for such quench experiments over equilibrium studies very close to the critical point, where the timescales become longer and the equilibration condition harder to fulfill.  Instead, in a quench scenario, the transition is crossed very rapidly from well outside its boundaries, and if the system has no time to respond to the change, then a full panoply of dynamical behavior can be observed that is characteristic of the symmetries broken by the transition (see Figure \ref{fig:diagram}).  This includes i) instability of the initial state and formation of seeds of the ground state on the opposite side of the critical point, ii) rapid expansion of the seeds to macroscopic proportions, and iii) slow growth to steady-state.  Using this approach, we have recently been able to make sub-Hz level (picoKelvin) precision measurements on the transition boundary in an antiferromagnetic spinor Bose-Einstein condensate, although the system temperature was 400 nK \cite{vinit2017}, well above the energy scales of the phase transition itself.

%In the current work we use a quantum quench to explore singular physics--the coupling between spin and spatial dynamics near a quantum phase transition--without necessitating an overly close approach to the singularity.  Quantum quenches have emerged as a powerful tool in this regard \cite{langen2015}.  If the transition point is crossed very rapidly from well outside its boundaries, and if the system has no time to respond to the change, then a full panoply of dynamical behavior can be observed that is characteristic of the symmetries broken by the transition (see Figure \ref{fig:diagram}).  This includes i) instability of the initial state and formation of seeds of the ground state on the opposite side of the critical point, ii) rapid expansion of the seeds to macroscopic proportions, and iii) slow growth to steady-state.  

In the present work we extend our earlier studies on the $q =0$ quantum phase transition in antiferromagnetic spinor Bose-Einstein condensate (BEC).  We present measurements of Hanbury-Brown and Twiss correlations and study their decay length.  While our previous works have confirmed the Bogoliubov predictions for the instability rate \cite{bookjans2011,vinit2017}, here we provide many details--both experimental and theoretical--that were missing in the earlier works.  For instance, we use statistical analysis to elucidate the variety and wealth of modes that were observed to form in the experiment, and compare with Bogoliubov predictions for these modes.  Similar to the quench experiments of reference \cite{Kang2017}, we have observed that the maximally unstable modes are localized near the center of the cloud, where the density is highest.  We show that this result appears naturally from the real space calculation of the Bogoliubov eigenvectors, which are different from the momentum modes typical of a homogeneous BEC.  

The paper is organized as follows.  Section II provides background on the phase transition being studied, and describes the experimental method in detail.  Section III contains the experimental data on Hanbury-Brown and Twiss correlations.  Section IV explains the theoretical technique and section V compares theory with a number of experimental observations.  Section VI is a conclusion and outlook.

\section{Background}

\subsection{Hamiltonian}

Multi-component spinor BECs offer unique opportunities to perform dynamical studies due to a number of quantum phase transitions that can be accessed within the spin sector of the Hamiltonian \cite{sten98spin,chan04,sadler06,black07,liu09,klempt2009,kronjager2010,zhao2015,seo2015}.  Consider the Hamiltonian for spin $F = 1$ Bose-Einstein condensates (see  \cite{Stamper-Kurn2013} and references therein):

\begin{equation}  
%H_{sp} =  \frac{c_2}{2} n({\bf r}) \langle \hat{\bf F}\rangle ^2 + p ({\bf r}) \langle \hat{F}_z \rangle + q \langle \hat{F}_z^2 \rangle  
H_{sp} =  \frac{c_2}{2} n({\bf r}) \langle \hat{\bf F}\rangle ^2 + q \langle \hat{F}_z^2 \rangle  
\label{eq:one} 
\end{equation}
The first term is the energy of spin-dependent interactions.  The interaction coefficient $c_2 = \frac{4 \pi \hbar^2}{3 M}(a_2-a_0)$ arises from spin-changing collisions that can convert two $m_F= 0$ atoms into an $m_F = \pm 1$ pair and vice-versa, a process constrained by the conservation of angular momentum.  Here $M$ is the atomic mass, and $a_{2,0}$ are the scattering lengths for atom pairs whose total angular momentum $F_{tot}=2$ and $0$, respectively.  These intrinsic interactions are antiferromagnetic for $c_2>0$ or ferromagnetic for $c_2<0$.  In this work we consider a sodium BEC, for which $c_2 >0$.  The overall density profile $n({\bf r})$ is determined by the chemical potential $\mu$, which is larger in magnitude by a factor of $100$ compared with Eqn.\ (\ref{eq:one}) for typical values of $q$.  Thus, we have removed the spin-independent terms from the Hamiltonian, which are assumed to be a constant.  ${\bf \hat{F}},\hat{F}_z$ are the vector spin-1 operator and its $z$-projection, respectively.  Hereafter in this work, we write $m \equiv m_F$.

%It consists of 3 terms.  The first is the energy of spin-dependent interactions.  The interaction coefficient $c_2 = \frac{4 \pi \hbar^2}{3 M}(a_2-a_0)$ arises from spin-changing collisions that can convert two $m_F= 0$ atoms into an $m_F = \pm 1$ pair and vice-versa, a process constrained by the conservation of angular momentum.  Here $M$ is the atomic mass, and $a_{2,0}$ are the scattering lengths for atom pairs whose total angular momentum $F_{tot}=2$ and $0$, respectively.  These intrinsic interactions are antiferromagnetic for $c_2>0$ or ferromagnetic for $c_2<0$.  In this work we consider a sodium BEC, for which $c_2 >0$.  The overall density profile $n({\bf r})$ is determined by the chemical potential $\mu$, which is larger by a factor of 100 than the second term in Eqn.\ \ref{eq:one} for typical values of $q$.  Thus, we have removed the spin-independent terms from the Hamiltonian, which are assumed to be a constant.  ${\bf \hat{F}},\hat{F}_z$ are the vector spin-1 operator and its $z$-projection, respectively.  Hereafter in this work, we write $m \equiv m_F$.

The second term in Eqn.\ (\ref{eq:one}) is the quadratic Zeeman shift due to the external magnetic field, $q = \tilde{q} B_0^2 + q_M$.  $B_0$ is the magnetic field at the trap center, $\tilde{q} = 276 $ Hz/Gauss$^2$ is the coefficient of the quadratic Zeeman shift for sodium atoms, and $g_F = 1/2$ and $\mu_B$ are the Lande $g$-factor and Bohr magneton, respectively \cite{ueda2012review}.  In addition to the static magnetic field, we introduce an additional term $q_M$, which is the shift caused by a microwave magnetic field through the AC Zeeman effect on the $F = 1, m$ sublevels \cite{bookjans2011,vinit2013}.  The microwave field generates an additional shift $q_M<0$, allowing us to access the $q<0$ region of the phase diagram without having to change the static magnetic field.

For an antiferromagnetic spinor BEC prepared in an initial state with zero net magnetization $\langle F_z\rangle=0$, the ground state of the above Hamiltonian for $q > 0$ is a polar condensate consisting of a single component--the $m = 0$ spin projection that minimizes $\langle \hat{F}_z ^2 \rangle$.  For $q<0$ the ground state maximizes the same quantity through a superposition of two components $m = \pm 1$, a so-called antiferromagnetic phase \cite{kawaguchi2012}.  We study the effect of a quantum quench across the zero temperature quantum phase transition  at $q = 0$.  

%The Hamiltonian written above is defined by the competition between the intrinsic magnetic interactions and the coupling to the external magnetic field.  Effects of the linear Zeeman term were considered in previous work through deliberate application of a field gradient \cite{vinit2017}.  However, for all of the data presented in this manuscript we tuned the magnetic field gradient along the trap axis to be zero, and thus only the quadratic Zeeman term plays a role in the dynamics.    With this assumption, for an antiferromagnetic spinor BEC prepared in an initial state with zero net magnetization $\langle F_z\rangle=0$, the ground state for $q > 0$ is a polar condensate consisting of a single component--the $m = 0$ spin projection that minimizes $\langle \hat{F}_z ^2 \rangle$.  For $q<0$ the ground state maximizes the same quantity through a superposition of two components $m = \pm 1$, a so-called antiferromagnetic phase \cite{kawaguchi2012}.  The symmetry of the ground state therefore changes discontinuously at $q = 0$, defining a zero temperature quantum phase transition \cite{kawaguchi2012}.  
%
\subsection{Experimental Method}

Sodium Bose-Einstein condensates in a single focus optical trap  were prepared in the $m = 0$ state in a static magnetic field.  The protocol is described in earlier work \cite{bookjans2011}, but we include relevant details here.  The peak density $n_0 = 5 \times 10^{14} \rm{cm}^{-3}$ and axial Thomas-Fermi radius $R_x = $ 340 $\mu$m were measured to an accuracy of 5\%, from which we determined the peak spin-dependent interaction energy $c_2 n_0 = h \times 120$ Hz.  The axial and radial trapping frequencies were 7 and $470$ Hz, respectively, accurate to 10\%.  The radial Thomas-Fermi radius was $R_\perp = $ 5 $\mu$m, and thus the aspect ratio of the cigar-shaped cloud was $\approx$ 70:1.  The measured temperature was 400 nK, close to the chemical potential of 360 nK.  

Our experiment required precise control over the bias magnetic field.  We applied a magnetic field $B_x$ aligned with the long axis $x$ of the cigar and tuned the field gradient $dB_x/dx$ to 
cancel ambient field inhomogeneities in our vacuum chamber to within $\pm$10 $\mu$G using a procedure that is described in \cite{vinit2017}.  The microwave field used to control $q_M$ was generated by an HP8648B synthesizer with $\sim 100$ Hz accuracy.  Its frequency was tuned below the ``clock'' transition, $|F = 1, m = 0 \rangle \rightarrow | F=2,m = 0 \rangle $ at 1.772 GHz \cite{kase89} by an amount between 260 to 470 kHz, resulting in a negative shift $q_M<0$ that counteracts the positive shift from the static field $\tilde{q} B_0^2$.   

The axial extent of the cloud, $R_x = 340 \mu$m, was substantially larger than the typical domain size of $\sim 30 \mu$m.  Thus multiple domains can form in this system.  Although the system is multi-mode, the spin modes are one-dimensional due to the quench energetics, as can be seen from the following argument.  The tight confinement along the transverse dimensions of the cigar implies a large transverse zero point energy for spin excitations.  We can estimate these from a two-dimensional circular box model similar to Reference \cite{scherer2010}.  The single particle eigenfunctions in transverse polar coordinates $\rho,\phi$ are $\propto J_l(\beta_{nl} \frac{\rho}{R_\perp})e^{il\phi}$, where $J_l$ is the $l$-th Bessel function with zeros $\beta_{nl}$.  The box eigenenergies are then $\epsilon_{nl} = \hbar^2\beta_{nl}^2/(2MR_\perp^2)$.  The lowest 3 energies are $\epsilon_{10},\epsilon_{11},\epsilon_{21}$ whose numerical values are, for our parameters, $h \times$ 50,130 and 230 Hz, respectively.  For the data collected here very close to the phase transition point, the quadratic Zeeman energy released was $h \times 8 $ Hz $ \ll \epsilon_{10}$ and thus transverse excitation was impossible.

\begin{figure}[htbp]
\begin{center}
\includegraphics[width=  \columnwidth]{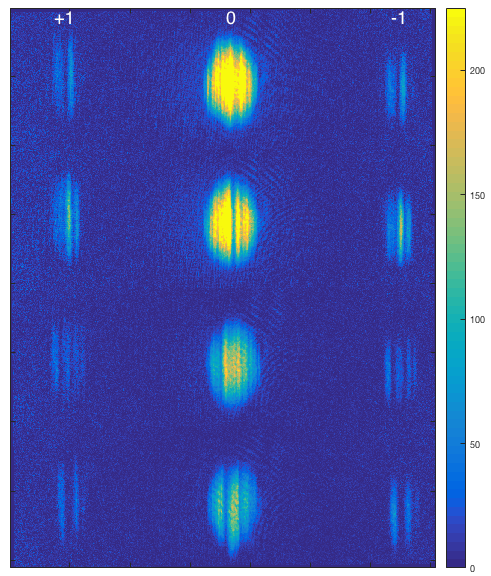}
%\vspace{-0.5in}
\includegraphics[width= \columnwidth]{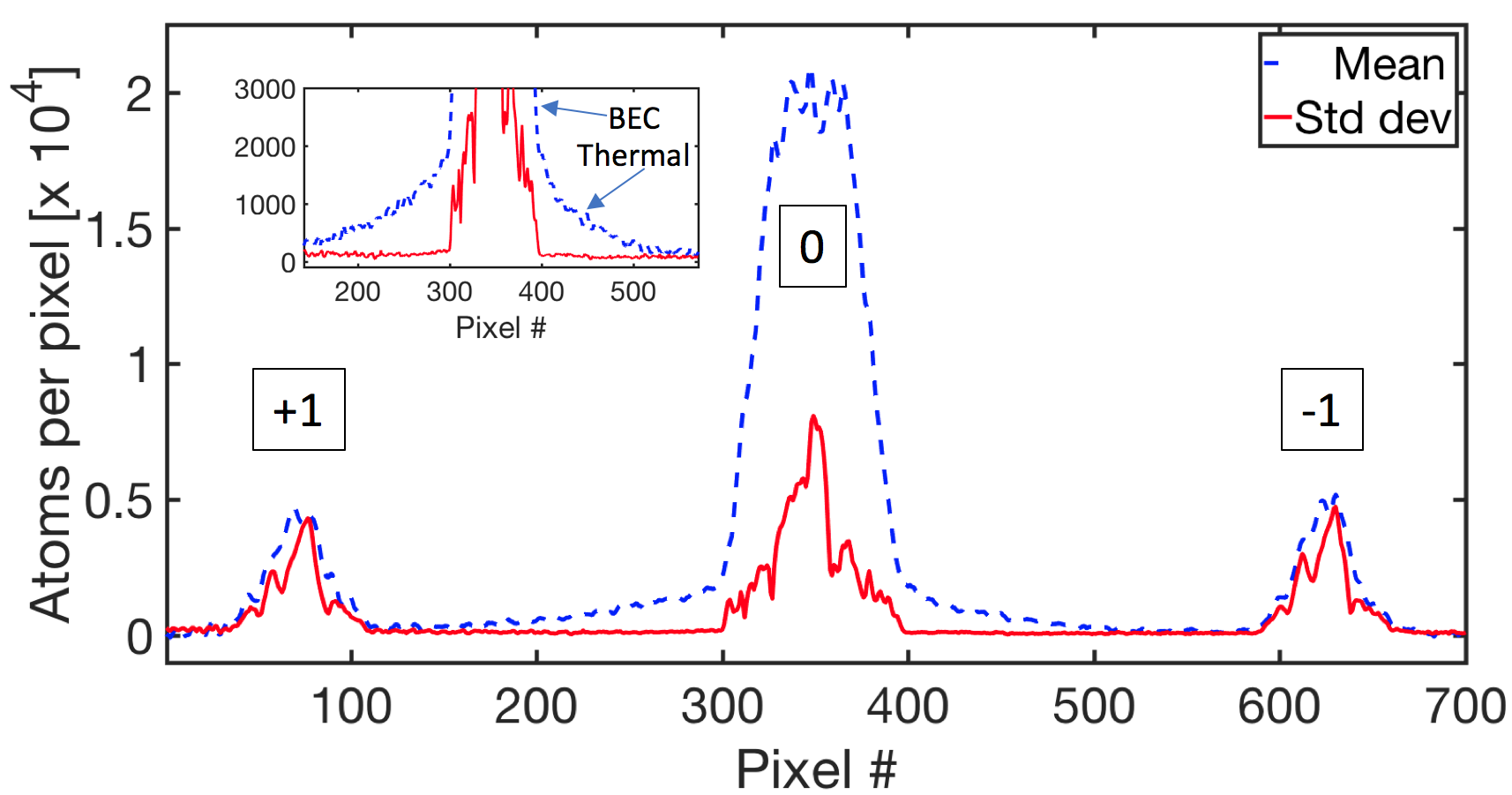}
\end{center}
\caption{(Color Online).  Stochastic nature of the quench.  Upper panel shows time-of-flight Stern-Gerlach images of 4 distinct, but otherwise identically prepared quenched Bose-Einstein condensates after a hold time of 20 ms.  Each image is 1.3 x 4.6 mm in size.  Lower panel shows mean and standard deviation of one-dimensional slices through the images.  The inset is a blow-up of the $m = 0$ condensate.}
\label{fig:SG_slices}
\end{figure}

The quench experiment consisted of rapidly switching $q$ from $q_i > 0$ to a final value $q_f< 0$ at $t=0$.  This was accomplished by sudden turn-on of the microwave field amplitude using a radiofrequency switch with sub-microsecond rise time.  Following a variable hold time, we switched off the trap and used time-of-flight Stern-Gerlach (TOF-SG) observations  \cite{Stamper-Kurn2013} to record an absorption image for each experimental run.  Due to the highly anisotropic expansion of the 70:1 aspect ratio, cigar-shaped BEC, the $x$-axis of the image was identical to the axial coordinate within the trap, with one minor difference--each spin component was offset by a fixed amount, $\propto m$, that was determined by its Stern-Gerlach separation in time-of-flight.  Using this information we could undo the SG separation to extract a one-dimensional density distribution in each of the 3 spin components, $n_i(x); i = 0,\pm 1$, with a spatial resolution of 10 $\mu$m \cite{vinit2013}.  

Time-of-flight Stern-Gerlach imaging is a highly sensitive technique, but can be limited by technical, rather than fundamental noise.  In our case, technical noise was mostly caused by interference fringes whose position varied from shot to shot.  We used powerful post-processing techniques to minimize this effect \cite{Ockeloen2010}.  This allowed us to observe data at hold times as short as 16 ms after the quench, where the time-of-flight $m = \pm 1$ atom signal on the camera corresponded to a mere 9 atoms/pixel$^2$.  The standard deviation of the resulting image noise could be reduced to a value that was no more than 1.7 times higher than the optical shot noise.  The sensitivity could not be further reduced without increasing the fluence of the imaging light to reduce shot noise contributions.  However, we found this incurred nonlinear effects and motion induced blurring of the cloud.  Future experiments aimed at increasing sensitivity could boost the atom signal at even earlier times by mixing the $m = \pm 1$ spins with that of the strong $m = 0$ cloud through spin rotations, effectively deploying a heterodyne technique.

%At $t = 16$ ms, the $m = +/-1$ atom number density detected was only 9 atoms per camera pixel on average, while the camera noise expressed in the same units was 15 atoms/pixel, yielding a signal-to-noise ratio of 0.6 for individual pixels.  The total camera noise was a factor 1.7 above the photon shot noise limit.  

\section{Experimental Results on Hanbury Brown-Twiss Correlations}

\subsection { Spin fluctuations}

Figure \ref{fig:SG_slices} encapsulates the multi-mode character of the instability, and reveals several interesting features.  The upper panel shows 4 representative Stern-Gerlach images taken on different experimental runs with a hold time of 20 ms after the quench.  Spin-exchange collisions are responsible for converting on average 15\% of the atoms from $m = 0$ into $m = \pm 1$ pairs, which form localized domains, seen as vertical stripes in the images.  Since the pairs were spin-correlated, the location and number of domains was highly correlated between $m = 1$ and $m = -1$ atoms, as also observed in earlier work \cite{vinit2013}.  In the current work we focus on the behavior within each spin cloud $m = \pm 1$, where the number and location of domains  varied stochastically from run to run.  This fluctuation is not due to technical reasons, but is a fundamental feature of the quench, as unoccupied spin excitation modes become rapidly occupied, with a highly variable number and mode distribution.  

We analyzed these fluctuations using statistical methods on an ensemble of measurements on identically prepared BECs.  Thirty images were collected at each quench hold time, whose mean atom number $\bar{N}$ and standard deviation $\sigma_N$ were computed.  We suppressed shot to shot atom number fluctuations by filtering only those whose atom number lay inside the range $\bar{N}<N<\bar{N}+\sigma_N$, resulting in samples with less than $10$\% atom number fluctuations.  This allowed us to more clearly observe the intrinsic noise.  

For this filtered data set, each camera image was reduced to a one-dimensional slice $f(x)$, where $x$ is the axial coordinate within the image.  To accomplish this, we converted the absorption images to atomic column density (atoms/pixel$^2$) through the known absorption cross-section for light resonant with the $F = 2 \rightarrow F' = 3$ transition \cite{kett99var,metc99}.  We then summed each image over the central 75\% of the time-of-flight Thomas-Fermi distribution along the radial direction.  The domain of the sum was chosen to maximize the number of atoms counted without introducing excess noise from the edges of the Thomas-Fermi distribution where the atom numbers were smaller than image noise.  The resulting set of slices ${f(x)}$, calibrated in atoms/pixel, were processed to derive mean and standard deviations.  The lower panel of Figure \ref{fig:SG_slices} shows these as dashed blue and solid red lines, respectively.  It is clearly visible in the figure that the $m = 0$ number density fluctuations were no more than 40\% of the mean, while those of the $m = \pm 1$ cloud were as much as 100\% of the mean.  As we will show, the two noise sources in fact have the same origin.

These results are intriguing, since a Bose-Einstein condensate is not expected to have 100\% variability in atom number.  Similar to laser light, a BEC can be described by a coherent state, with Poissonian number density fluctuations (shot noise).  In the absence of any quench, an analysis of individual pixels showed that the $m = 0$ cloud possessed a variance quite close to this limit.  For shot noise, the ratio of standard deviation to mean atom number density $\langle \Delta n \rangle/\bar{n} = 1/\sqrt{\bar{n}}$ is very small for the large atom numbers present in the experiment.  By contrast, the  $m = \pm 1$ clouds show ``super-Poissonian'' (SP) noise with $\langle \Delta n \rangle/\bar{n} = 1$.  We note that SP fluctuations have been observed in related works where spin-exchange collisions are responsible for the statistical fluctuations \cite{Lucke2011,Gross2011}.  

\subsection{Role of thermal atoms in the quench}
%{\bf Thermal atoms}. 
Our data in the inset to the lower panel of Figure \ref{fig:SG_slices} also reveal that thermal atoms play a negligible role in the quench.  The  $m = 0$ cloud can be seen to possess a bimodal distribution in space, with the central, sharp peak corresponding to the Bose-Einstein condensate and a lower density, more diffuse pedestal due to the thermal atoms whose occupation fraction was 40\%, corresponding to a temperature of $T \sim 400$ nK.  

In spite of such a large thermal population, at short hold times the $m = 0$ super-Poissonian spin fluctuations clearly occur only in the condensate, and not in the normal gas, as the fluctuations (the red curve) drop sharply to zero at the Thomas-Fermi radius.  At longer hold times $t = 48$ ms (not shown in the figure), the $m = \pm 1$ populations had grown substantially to comprise a fraction $0.6$ of the total condensate.  By this time, a thermal gas of $m = \pm 1$ atoms had become populated via interactions between the thermal cloud in $m = 0$ and various condensed spin components \cite{Schmaljohann2004}.  Nonetheless, the $m = -1$ spin fluctuations (and therefore, spin relaxation dynamics) still continued to occur only in the condensate, and not in the normal gas.  

Our results can be summarized as coherent spin evolution for short times, followed by thermally assisted spin redistribution occurring at long hold times.  The separation of timescales poses the intriguing possibility to cool the sample via spin-changing collisions and selective spin state removal of the thermal cloud.  

\subsection{Hanbury Brown-Twiss correlations}
%\noindent{\bf Hanbury Brown-Twiss correlations}.  
Turning now to the condensed components, the super-Poissonian fluctuations mentioned earlier lead to Hanbury Brown-Twiss correlations in our spatially extended one-dimensional system.  Such correlations have been widely observed from the domain of radio-frequencies where they were initially observed and used to determine the angular diameter of stars \cite{HANBURYBROWN1956} to the optical domain \cite{loud83}.  For massive particles, these correlations have been observed with ensembles of ultracold atoms \cite{Esteve2006,Jeltes2007,Perrin2012}, as well as in the sub-atomic realm, where one can use them to extract information about nuclear structure from collisions \cite{Weiner2000}.  In our experiment with Bose-Einstein condensates they reveal the coherence length associated with the non-equilibrium state created by the quench.  

We examined the second order spin density correlation function $g_2(x)$ with relative coordinate $x$ between points of observation:  
\begin{equation}
g_2(x) = \left \langle \frac{\langle n(x_0+x) n(x_0) \rangle}{\langle n(x_0+x)\rangle \langle n(x_0)\rangle} \right \rangle_{x_0} 
\end{equation}
Here $n(x_0)$ is the density of a particular spin component, for example, the $m = -1$ atoms, and $x_0$ represents a spatial position within the cloud.  The inner and outer brackets refer to ensemble and spatial averages, respectively.  Since the Thomas-Fermi density profile breaks translation invariance in the system, we define ensemble and spatial averages to be distinct quantities.  First ensemble averages were taken over a restricted set of images with reduced atom number fluctuations as discussed earlier, after which a spatially averaged $g_2$ was computed over all values of $x_0$.  For each data set we performed bimodal fits to the central, $m = 0$ cloud to determine the condensate Thomas-Fermi radius $R_{TF}$.  From the mean of the $m = \pm 1$ slices we computed the central position $x_c$ of the $m = \pm 1$ clouds in order to undo the Stern-Gerlach expansion.  From $x_c$  and $R_{TF}$ we could generate distributions $n_{-1}(x)$ within the cloud.  Since one divides by the mean value to compute $g_2$, its value blows up as one approaches the distribution edges--to avoid this, we restricted our analysis to $-R_{TF}/2 < x < +R_{TF}/2$ and used periodic boundary conditions to compute the averages over $x_0$ \footnote{Technically, $g_2(0) = \langle n(n-1)\rangle/\langle n \rangle^2$, but we can neglect the difference between $n$ and $n-1$ for our experiment, where macroscopic numbers of atoms are detected.}.

In optics, the equivalent of the spatial variable $x$ is the delay time $\tau$ between light paths in an optical system.  In this realm, a pure coherent state has Poisson fluctuations which result in $g_2(x) = 1$ for all values of $x$ \cite{loud83}.  Excess fluctuations are evidence of ``photon bunching''--a tendency of photons to arrive at the same time due to their bosonic nature, resulting in $g_2(0)>1$.  Thermal, or chaotic light fields such as are emitted by a lamp or other low coherence source, possess $g_2(0) = 2$, with $g_2(x) \rightarrow 1$ as the delay time exceeds the coherence time.

\begin{figure}
\includegraphics[width=\columnwidth]{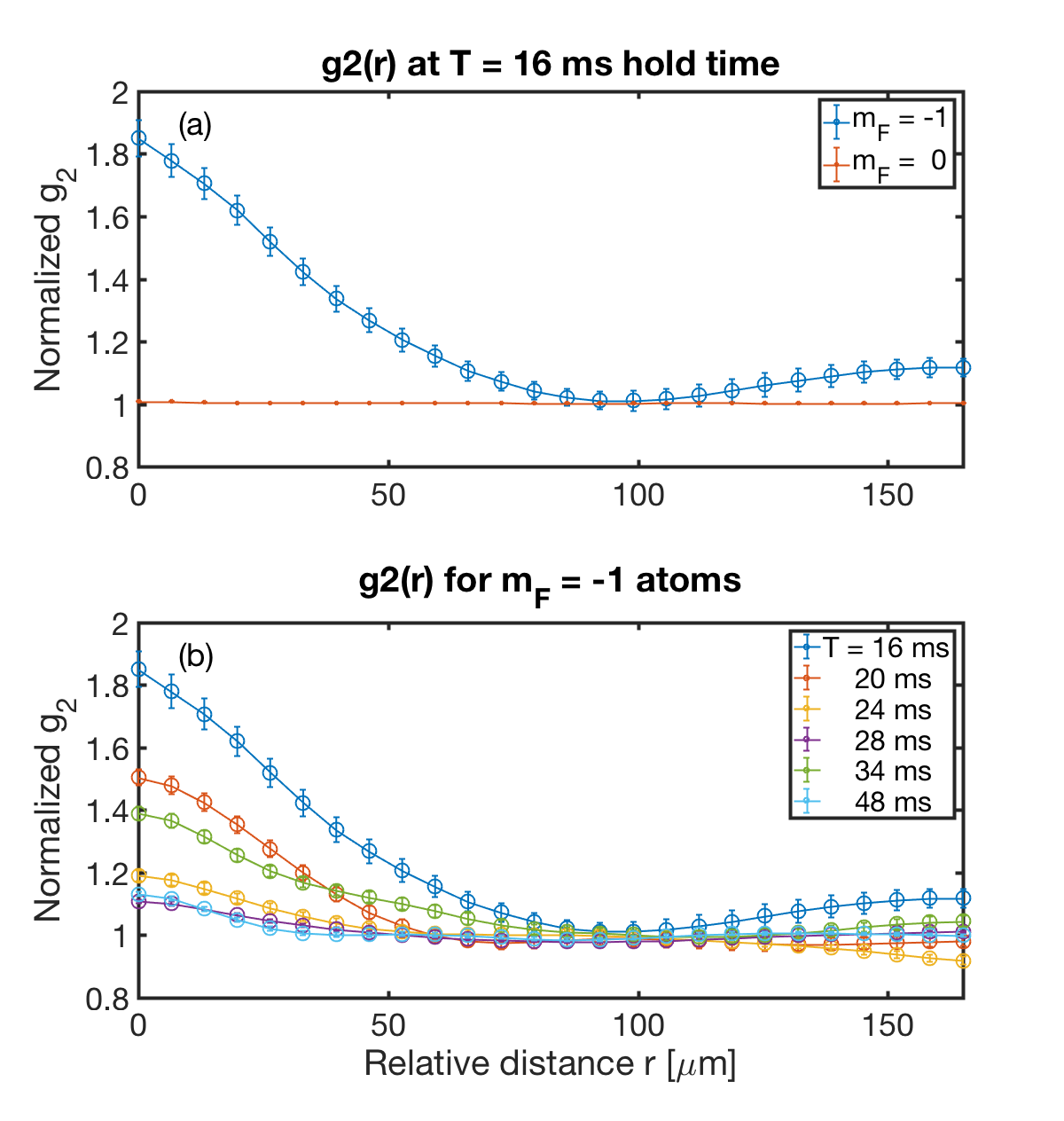}
\includegraphics[width=\columnwidth]{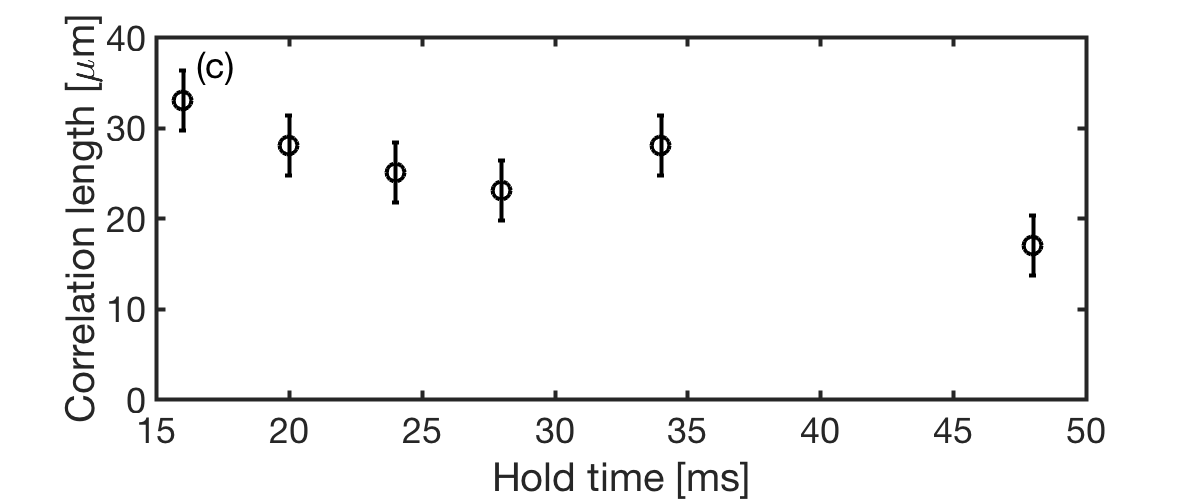}
\caption{(Color Online).  Hanbury Brown-Twiss correlations.  (a) and (b) Second order correlation functions $g_2(x)$ versus separation $x$, for parameters defined in the text. Error bars are statistical.  (c) Correlation length defined as the half--width of $g_2-1$.  Error bars are the experimental resolution.}
\label{fig:sp_noise}
\end{figure}

Our experiment clearly demonstrates the correspondence with thermal light, although, as shown earlier, the thermal cloud itself plays no role in the instability.  In Figure \ref{fig:sp_noise}a we have evaluated the normalized $g_2(x)$ for our experimental data using the $m = -1$ and $m = 0$  atom distributions at a time $16$ ms after the quench.   Corresponding data for $m = +1$ was similar to $-1$. The data clearly show a very strong ``spin bunching'' effect, as individual $m = -1$ atomic spins tend to be co-located, with $g_2(0) = 1.9 > 1$.  This was observed very early in the quench, when the $m = +/-1$ atom numbers were very small, and the corresponding fluctuations proportionally large.  The measured value of $g_2(0)$ close to $2$ is consistent with a thermal state and the super-Poissonian noise in the population. 

In contrast to the above bunching phenomenon, the $m = 0$ spins exhibited very little bunching--the measured $g_2(x)$ was very close to 1 for all values of $x$, consistent with a Bose-Einstein condensate in a coherent state.  Upon closer examination, we determined that $g_2(0)-1 \approx 0.01$.  This small excess in the normalized variance can be largely explained by technical noise caused by atom number fluctuations as well as spin-exchange noise in the $m_F = 0$ population due to the production of $m = \pm 1$.  The former (latter) had a standard deviation of $7\%$ ($5\%$), and were thus of the same magnitude at this early hold time.  For only slightly later times of $t \ge 20$ ms, the stochastic fluctuations of the quench became $40 \%$ as noted earlier in Figure \ref{fig:SG_slices}, and dominated over technical noise.
  
%This tiny excess can be attributed to the correlated super-Poissonian fluctuations of the $\sim 5\%$ $m = \pm 1$ atom fraction, as well as the $7\%$ total atom number fluctuations.  
%At this early hold time, these contributions were of the same magnitude, whereas at only a slightly later time of $20$ms, the stochastic fluctuations of the quench dominated over technical noise.  

For the $m = -1$ atoms, Figure \ref{fig:sp_noise}b shows that $g_2(x)$ rapidly decays in space, exhibiting damped oscillations that approach a value of 1.   Defining the correlation length, $x_{1/2}$, through the formula $\frac{g_2(x_{1/2})-1}{g_2(0)-1} = 0.5$ we obtained $x_{1/2} = 33 \mu$m.  This correlation length is substantially smaller than the condensate Thomas-Fermi radius $R_{TF} = 340 \mu$m, and reveals the average spatial extent of the spin modes excited by the quench.  The oscillations in the $g_2$ function indicated the formation of multiple domains simultaneously, and is further evidence of the multi-mode character that we explore in the next section.  Individual images revealed domains between $15-45 \mu$m in size (see Figure \ref{fig:unstable_modes1}).  The middle panel of Figure \ref{fig:sp_noise} shows that for slightly longer hold times where the ratio of $m = \pm 1$ to $m = 0$ populations became appreciable, $g_2(0)\rightarrow 1$ indicating the formation of a more stable, non-equilibrium state.  In spite of this, the population in $\pm1$ had not yet reached its equilibrium value.  The lower panel of Figure \ref{fig:sp_noise} shows that the correlation length shrinks by a factor of about 2 with hold time (although an oscillation in the population seen in Figure \ref{fig:diagram} causes it to momentarily increase at 34 ms).  Thus the system transferred energy from long to short wavelength modes as the quench progressed.

\begin{figure*}[htbp]
\begin{center}$
\begin{array}{ccc}
\hspace{-0.1in}
\includegraphics[width= 0.7\columnwidth]{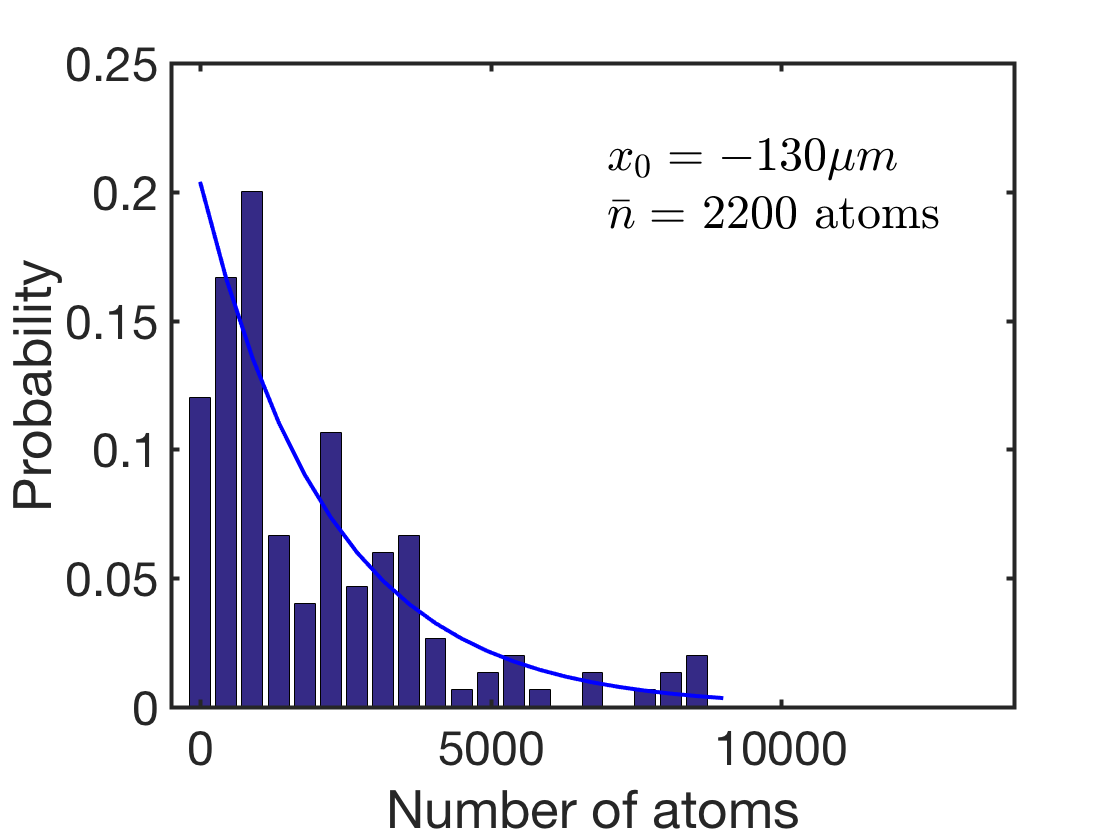}&
%\hspace{-0.25in}
\includegraphics[width= 0.7\columnwidth]{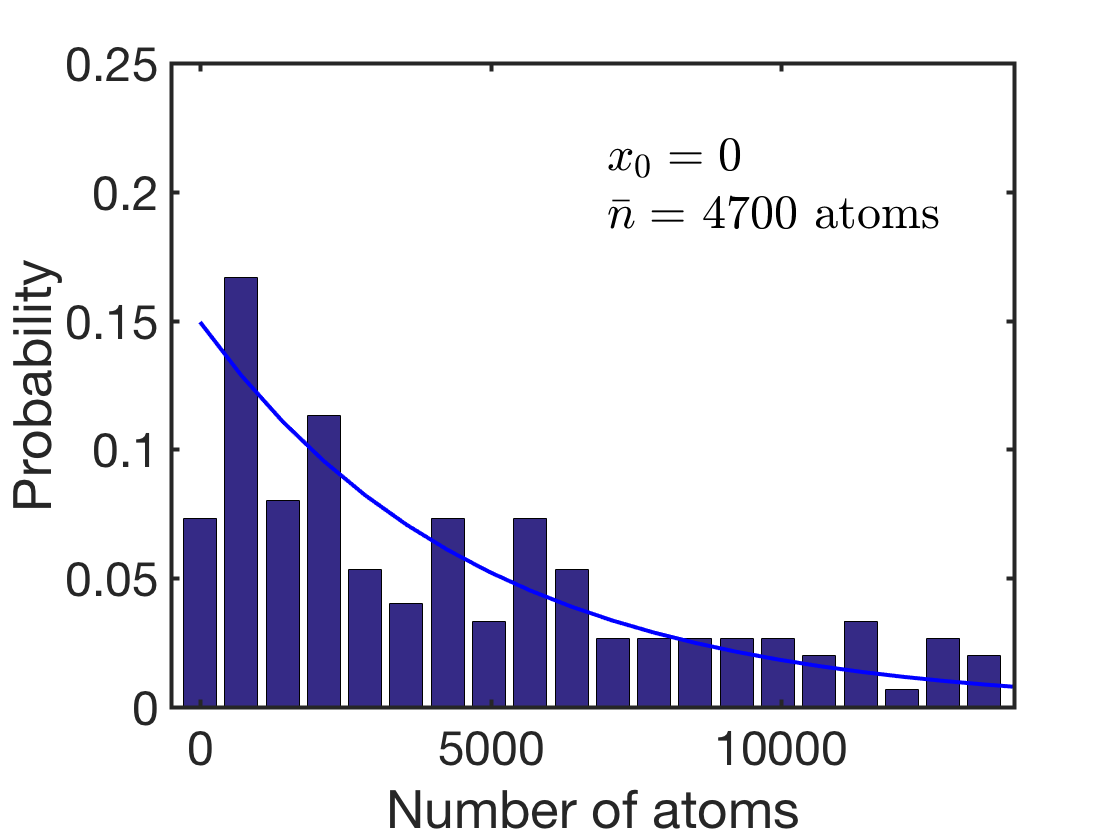}
%\hspace{-0.25in}
\includegraphics[width= 0.7\columnwidth]{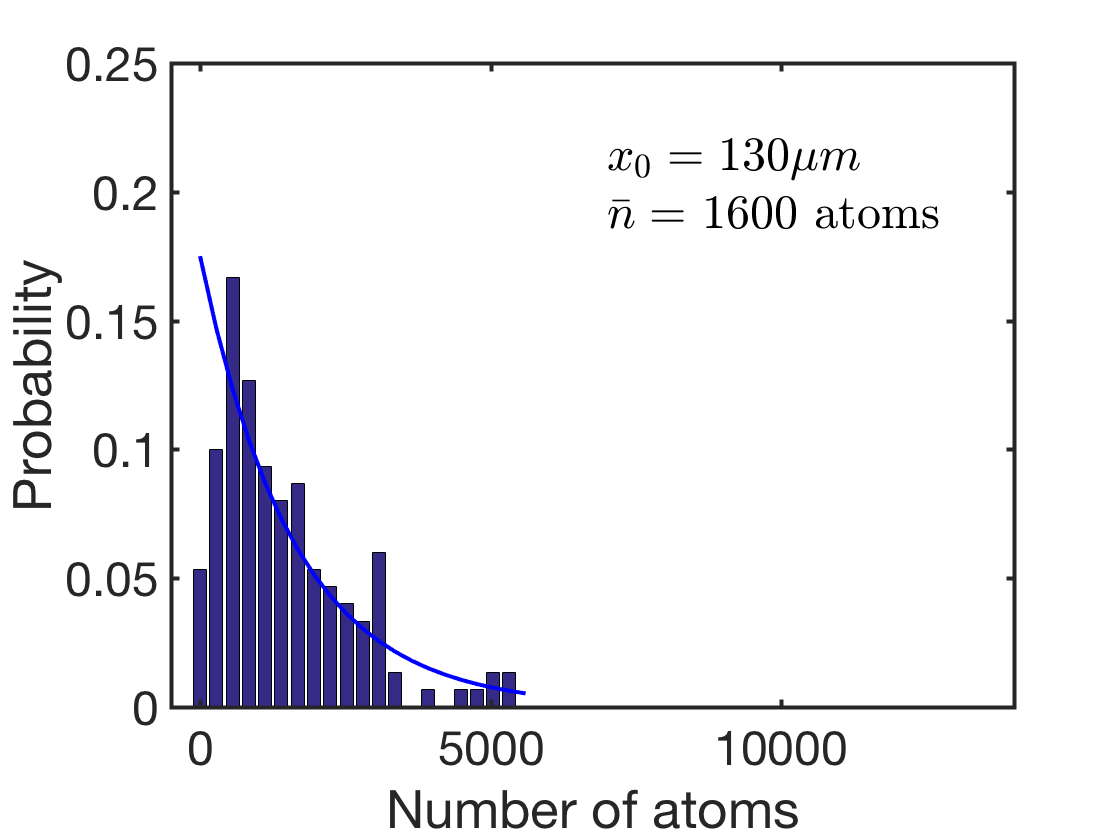}
\end{array}$
\end{center}
\caption{(Color Online).  Quantum spin thermalization.  Shown are the probability distributions for $m = -1$ atoms at a hold time of 20 ms, at different spatial locations $x_0$ within the cloud that have varying mean atom number $\bar{n}(x_0)$.  Each graph uses 20 bins  covering the region of non-zero data, approximately $0<n<3\bar{n}(x_0)$. Good agreement is found with the Bose-Einstein distribution (Eqn.\ \protect{\ref{eq:mias}}, solid curves) which uses the local sample average $\bar{n}(x_0)$ but otherwise has no fitting parameters.}
\label{fig:probability}
\end{figure*}

\subsection{Boltzmann statistics for the spin}
%\noindent {\bf Thermal Statistics}. 
How does a thermal state appear within a Bose-Einstein condensate, whose dynamics are governed by quantum mechanics?  A closely related, and more general question is how isolated quantum systems thermalize when placed out of equilibrium \cite{langen2015}.  One answer that has emerged in recent years addresses the similarity between quantum and thermal fluctuations, particularly when one looks at the system locally.  That is, even though the entire system is quantum, and has a pure case density matrix $\rho$, for which Tr$(\rho) = 1$, a subsystem $A$ will have a reduced density matrix $\rho_A = $Tr$_{\bar{A}}(\rho)$ that is mixed.  Here, the trace is taken over $\bar{A}$, the part of the system not in $A$.  Measurements made within A will be indistinguishable from those made on a global thermal ensemble, since the entanglement that is generated between $A$ and the rest of the system by the quench is not detected.  This notion has recently been tested in site-resolved optical lattices, where the sub-system consists of a finite number of sites \cite{Kaufman2016}.

In our case ``local'' refers to a sector within the spin space of all particles.  In particular, since the $m=0$ atom number is $10^6-10^7$, these atoms behave classically.  The quantum behavior is restricted to the $|m| = 1$ sector, within which there is entanglement between $+1$ and $-1$ spins generated by the quench.  A measurement of both spins together shows strong quantum correlations, and has been observed previously \cite{bookjans2011b,Lucke2011,Gross2011,vinit2013}.  Measurement of a subspace consisting of just one of the spins should result in a mixed case density matrix.  Thus our experiment realizes quantum thermalization within spin space, analogous to the real space thermalization of Kaufman et al.\ \cite{Kaufman2016}.  

A theoretical prediction of Mias et al.\  \cite{Mias2008} elaborates upon this idea.  Using a Bogoliubov treatment they showed that in a quench experiment, if one observes either of the two spin states, $+1$ or $-1$, the result is a Bose-Einstein probability distribution for the number of atoms $n_k$ in the $k$-th spatial mode,
\begin{equation} 
P(n_k) = \frac{1}{\bar{n}_k+1} \left ( \frac{\bar{n}_k}{\bar{n}_k+1} \right )^{n_k} \approx e^{-n_k/\bar{n}_k}
\label{eq:mias}
\end{equation}
where $\bar{n}_k$ is the mean number of atoms in that mode, a number that grows exponentially with time subsequent to the quench.  In the above formula the latter approximation holds for $\bar{n}_k \gg 1$, which holds for all of our experimental data.  From Eqn.\ \ref{eq:mias}, we can see that the distribution of just one of the spins should obey Boltzmann statistics with an effective temperature $T \propto \bar{n}_k$.  

In Figure \ref{fig:probability} we make a direct experimental comparison with the predicted probability distribution, Eqn.\ \ref{eq:mias}, at a hold time of $20$ ms after the quench.  Shown are probability histograms for the number of detected $m = -1$ atoms at 3 different spatial locations within the cloud, $x_0 = -130 \mu m, 0$ and $+130 \mu m$, where the mean atom number, $\bar{n}$, varied due to the inhomogeneous Thomas-Fermi density distribution.  We used different spatial locations $x_0$ spaced by much more than the correlation length of Figure \ref{fig:sp_noise}c, in order to demonstrate that the probability distribution is a {\em local} quantity, and varies throughout the cloud.  The theoretical prediction from Eqn.\ \ref{eq:mias} is plotted as a solid line.  It uses this sample mean as its only adjustable parameter.  The agreement between the data and theory in each case is quite good.  To generate sufficient statistics to generate an entire probability distribution from our limited data set, we used a 15 pixel ($100 \mu$m) wide sample centered at $x = x_0$, and 10 experimental runs whose atom number fluctuations had been filtered to $<5 \%$, as discussed earlier.  Thus each graph had 150 data points, and the mean atom number $\bar{n}(x)$ was evaluated at $x=x_0$.  By averaging over spatial pixels we necessarily included data with different values of $\bar{n}$, by $\pm 50$\% for $x_0 = \pm 130 \mu m$ and $\pm 10$\% for $x_0 = 0$.  In spite of this, the local exponential character of the distribution clearly persists, and reveals the thermal statistics of the spin states produced by the quench.

\section{Theory}

Having directly generated the modes experimentally, we turn now to their theoretical description.  We focus our theory on both $q <  0$ and $q > 0$ with zero magnetization.  Our effort closely parallels that of other experimental observations of spinor instabilities, with important differences.  For example, Bogoliubov theory was applied to a finite $q>0$ instability of ferromagnetic $F = 1$, $^{87}$Rb spinor BEC \cite{sadler06}, as well as to the $q=0$ instability \cite{scherer2010} and other instabilities \cite{kronjager2010} of antiferromagnetic $F = 2$ spinor BEC.  Broadly speaking, these works have identified instabilities arising either through bulk modes with a finite wavevector, as in \cite{sadler06,kronjager2010}, or a specific mode or set of modes that are resonantly excited at certain values of the quadratic Zeeman tuning parameter \cite{scherer2010}.  Our studies, by contrast, explore an intermediate regime.  A bulk analysis assuming spatial homogeneity fails to capture essential features of our observations, particularly the localized instability near the trap center.  However, neither is our experiment dominated by the discrete mode structure of the trap, as the relevant modes along the long axis of the cigar are too closely spaced for us to resolve.  Instead, in our specific experimental geometry, the inhomogeneous density profile plays an important role in shaping the unstable modes.  We uncover these modes by solving the Bogoliubov equations directly in coordinate space.

Bogoliubov theory  was first applied to multicomponent (spinor) BEC separately by Ho \cite{ho98} and Ohmi and Machida \cite{ohmi98}.   Following their approach and others \cite{kawaguchi2012}, one linearizes the spinor Gross-Pitaevskii (GP) equations (or the corresponding Heisenberg equations of motion for the field operators) about an initial state that is classical.  In our case and several of the examples above, this is a state $\psi_0$ consisting of all atoms in the $m = 0$ sublevel, with only small corrections $\delta \psi_{\pm 1}$ describing the populations in $m = +1$ and $-1$.   Due to the small ratio $c_2/c_0$, we assume that the spin instabilities do not couple strongly to density fluctuations, and thus we can neglect fluctuations in the $m = 0$ state.  The resulting spinor wavefunction may be written as 
\[ \Psi = \Psi_0 +\delta \Psi = \left( \begin{array}{c}0\\ \psi_0 \\ 0 \end{array} \right ) + \left( \begin{array}{c}\delta \psi_{+1} \\ 0 \\ \delta \psi_{-1} \end{array} \right )  \]
and the resulting linearized spinor GP equation for $m = \pm 1$ is \cite{saito05}:
\[i \hbar \frac{\partial \psi_m}{\partial t}  =  \left ( -\frac{\hbar^2}{2 M} \nabla^2 - p(x) m  + q m^2 \right ) \psi_m \]
\begin{equation}
+ U(x) \left ( \psi_m + \psi_{-m}^* \right )  
\label{eq:GPlin}
\end{equation}
In the above we have simplified the notation to $\psi_m \equiv \delta \psi_m$, and assumed a one-dimensional description with axial coordinate $x$, so that $U(x) = c_2 n_0(x) = c_2 |\psi_0(x)|^2$.  To model the data presented in this paper we set the linear Zeeman term $p(x) = 0$ and allow the quadratic Zeeman shift $q$ to vary as a free parameter.  We exclusively study the regime very close to the phase transition, i.e., $-U_0 \ll q < 0$, where $U_0 = c_2 n_0(x=0)$ and $n_0(x=0)$ is the peak $m = 0$ atom number density at the trap center.

We expand the wavefunctions in a basis of spin excitations, with spatial mode index $k$ and frequency $\omega_k$:
\begin{equation}
\psi_{m} (x,t) = \sum_k u_{k,m} (x) e^{-i \omega_k t} + v_{k,m}^*(x) e^{+i \omega_k t} 
\label{eq:expansion}
\end{equation}
Note that since the above is actually two equations, one each for $m = \pm 1$, there will be two spin modes associated with each spatial mode $k$.  Putting Eqn.\ (\ref{eq:expansion}) into Eqn.\ (\ref{eq:GPlin}) and equating terms with equal time-dependence, we get the Bogoliubov-de Gennes equations:
\[E u  =  \left (- \frac{\hbar^2}{2 M}  \nabla^2  + q -p(x) +U \right ) u + U v \] 
\begin{equation}
- E v  =  \left (- \frac{\hbar^2}{2 M}  \nabla^2  + q +p(x) +U \right ) v + U u 
\label{eq:bog}
\end{equation}
where
\[ \left ( \begin{array}{c} u_{k,1} = u \\ v_{k,-1}= v \\ E \end{array} \right ) {\rm and}  \left ( \begin{array}{c} u_{k,-1} = v \\ v_{k,1} = u \\ -E \end{array} \right )\]
with $U \equiv U(x)$.  The energy eigenvalues of the spin wave modes are $E \equiv E_k$.  They are collective excitations of the spin degrees of freedom about the $m = 0$ condensate.  Due to the conservation of spin, each of these quasiparticles is comprised of atom pairs of equal and opposite spin projection $m = \pm 1$.  For an anti-ferromagnetic spinor BEC, where $c_2 > 0$, the $m = \pm 1$ components experience a repulsive interaction with the $m = 0$ condensate.  These excitations are very similar to the density modulations of a single component BEC, where excitations are comprised of atom pairs with equal and opposite momenta $\pm {\bf k}$, and which are repelled from the ${\bf k}=0$ condensate \cite{pita03book}.  For the spin wave case, however, the dynamics are much slower than for sound waves by the factor $\sqrt{c_2/c_0}$ ($\approx 8$ for sodium atoms), as first elucidated by Ho \cite{ho98} and Ohmi and Machida \cite{ohmi98}.  Thus the lowest excitation frequencies are typically in the range of a few Hz to 10s of Hz.

We find the eigenvalues and eigenvectors of Eqns.\ (\ref{eq:bog}) using a straightforward matrix diagonalization in MATLAB \cite{matlab}.  Figure \ref{fig:eigenvalues} shows the numerical solutions for the energy eigenvalues for some representative parameters.  For $q>0$ the eigenvalues are real.  If $q$ is chosen to be negative, one or more collective modes in Eqn.\ (\ref{eq:expansion}) has an energy eigenvalue $E_k$ which crosses into the complex plane.  This triggers an exponential growth in the population of those modes, which are linear combinations of the spin states $\pm 1$.  Thus the populations $ \psi_{m}^\dag \psi_{m} $, for $m = \pm 1$, also grow exponentially with time, similar to a parametric amplifier \cite{walls2008book}.  Although we have not written down the Bogoliubov expansion in terms of the field operators $\psi_m$, it is straightforward to do so, and all quantum effects and correlations can be calculated in a straightforward manner \cite{kawaguchi2012}.

Before turning to solutions to the equations, we point out some differences between the coordinate and momentum representations.  For uniform systems, Bogoliubov theory is best described in momentum space, using plane wave modes.  One can then write the annihilation operator for a boson with momentum ${\bf k}$ in terms of corresponding operators for quasiparticles with momenta ${\bf k}$ and ${\bf -k}$.  
The Bogoliubov transformation contains within it, therefore, a direct correlation between quasi-particles of opposite momenta.  This correlation is similar to that obtained in the Bogoliubov diagonalization of the Hamiltonian of a single component weakly interacting Bose gas \cite{peth02book}.  In addition, in the multicomponent Hamiltonian, Eqn.\ (\ref{eq:one}), the correlation is between quasiparticles of opposite spin.

In our experiment, where the $m = 0$ condensate has a Thomas-Fermi spatial density profile, an expansion in momentum eigenstates is not useful.  Instead, we have followed the approach of Ruprecht et al.\ in the analysis of collective excitations of a scalar BEC in a trap \cite{rupr96}.  In that case, the Thomas-Fermi density profile led to collective mode functions that were spatially varying, and which represented modes located inside of or near the Thomas-Fermi surface.  

We will find the same to be true of the collective spin modes for a spinor BEC under harmonic confinement.  An alternate way to view these modes is in terms of standing wave solutions $u_{k,m}(x),v_{k,m}(x)$ for the small excitations $\pm 1$ that are created within the Thomas-Fermi boundaries of the $m = 0$ cloud (see Figure \ref{fig:unstable_modes1}A for an example).  Thus, rather than momentum correlations, as expected for a uniform system, the Bogoliubov analysis reveals the {\em spatial} correlations for particles of opposite spin $m = \pm 1$, as noted in Figure \ref{fig:SG_slices} and our earlier work \cite{vinit2013}.  The correlations only exist, however, within the domains defined by those modes.

\subsection{Uniform density profile} 

The case of a uniform $m = 0$ density, $U = $ constant, is a useful point of reference since the solution can be analytically obtained.  The energy spectrum in this case is
\begin{equation} 
E_k =  \sqrt{(\epsilon_k+q)(\epsilon_k+q+2 U)} 
\label{eq:Ehom}
\end{equation}
where $\epsilon_k = \hbar^2k^2/(2M)$, $k =  \pi/L \times n$, $n = 1,2,3,...$, for excitations in a box of length $L = 2 R_{TF}$, where $R_{TF}$ is the axial Thomas-Fermi radius.  The Bogoliubov eigenfunctions are box modes 
\begin{equation} \phi_n(x) = \sqrt{\frac{2}{L}} \sin{\left [\frac{n \pi}{L}(x+L/2) \right ]}\label{eq:box} \end{equation}  
for $|x| < L/2$.
For our parameters the ground state energy of the box $\epsilon_1 \simeq h \times$ 0.05 Hz is smaller than our experimental resolution, so we can assume a quasi-continuous spectrum.  Thus for $q>0$ all eigenvalues are real, while for $q < 0$ imaginary eigenvalues define unstable modes.  For short times after the quench, the amplitude of these unstable modes grows exponentially in time with a rate $\Gamma = |{\rm Im}(E_k)|/\hbar$.  The {\em maximally unstable mode} is defined to be the one whose imaginary component is the largest, i.e., $\Gamma = \Gamma_{max}$.  For  $-U < q < 0$, maximizing $\Gamma$ yields the mode 
\begin{equation}
\phi_1(x) = \sqrt{\frac{2}{L}} \sin{\left [\frac{\pi}{L}(x+L/2) \right ]}
\label{eq:mode}
\end{equation} 
It has a wavevector $k_{max} = \pi/L$, i.e. a wavelength twice the Thomas-Fermi length of the condensate.  Neglecting the zero point energy, the corresponding instability rate is
\[ \Gamma_{max} \approx |\sqrt{q(q+2U)}| / h \]

\subsection{Thomas-Fermi density profile}

To solve the Bogoliubov Eqns.\ (\ref{eq:bog}) in the inhomogeneous case, we expand $u,v$ in the box basis (Eqn .\ (\ref{eq:box})), which for $p = 0$, yields the matrix equation
\[
\left ( \begin{array}{cc} H^0_{\mu \nu}  & U_{\mu \nu} \\ -U_{\mu \nu} & -H^0_{\mu \nu}\\ \end{array} \right ) 
\left ( \begin{array}{c} u_\nu \\ v_\nu \\ \end{array} \right ) 
= E \left ( \begin{array}{c} u_\nu \\ v_\nu \\ \end{array} \right) 
\] 
where the basis size was held to $N$ elements, $u_\nu,v_\nu$ are the box basis coefficients for $\nu = 1,2,3,...N$, and a summation over $\nu$ is implied in the matrix product.  In this basis, the matrix elements of the operators in Eqn.\ (\ref{eq:bog}) are $H^0_{\mu \nu} = (\epsilon_\mu+q)\delta_{\mu \nu}$ and $U_{\mu \nu} = U_0 \int_{-L/2}^{L/2} \phi_\mu (1-4x^2/L^2) \phi_\nu dx$.  The latter is easily computed,  
\[\frac{U_{\mu \nu}}{U_0} = \left \{ \begin{array}{ll}
-\frac{16 \mu \nu (1+(-1)^{\mu+\nu})}{\pi^2 (\mu^2-\nu^2)^2} & \mbox{if } \mu \neq \nu \\
\frac{2}{3}+\frac{2}{\pi^2 \mu^2} & \mbox{if } \mu = \nu
\end{array}
\right. 
\] 
The box eigenenergies are $\epsilon_\mu = \mu^2 \epsilon_1$.  

The numerical problem consisted of diagonalizing a square matrix of order $2N$, with $N$ pairs of eigenvalues $E,-E$ corresponding to the pair of spin modes discussed earlier.  For each value of $q$ and given the values of $\epsilon_1,U_0$, the eigenvalues and eigenvectors were found numerically using MATLAB.  The routines were tested against the exact solutions, Eqns.\ (\ref{eq:Ehom}) and (\ref{eq:box}), by fixing the density to be a constant.  Typically, the ground state energy was found to converge to $10^{-5}$ using 150 basis elements.  Unless otherwise stated, the parameters used were $U_0/h = 96$ Hz and $\epsilon_1/h = 0.0047$ Hz.
These values corresponded closely to typical experimental parameters.

\subsection{Eigenvalue Spectrum}

\begin{figure} [htbp]

\includegraphics[width= \columnwidth]{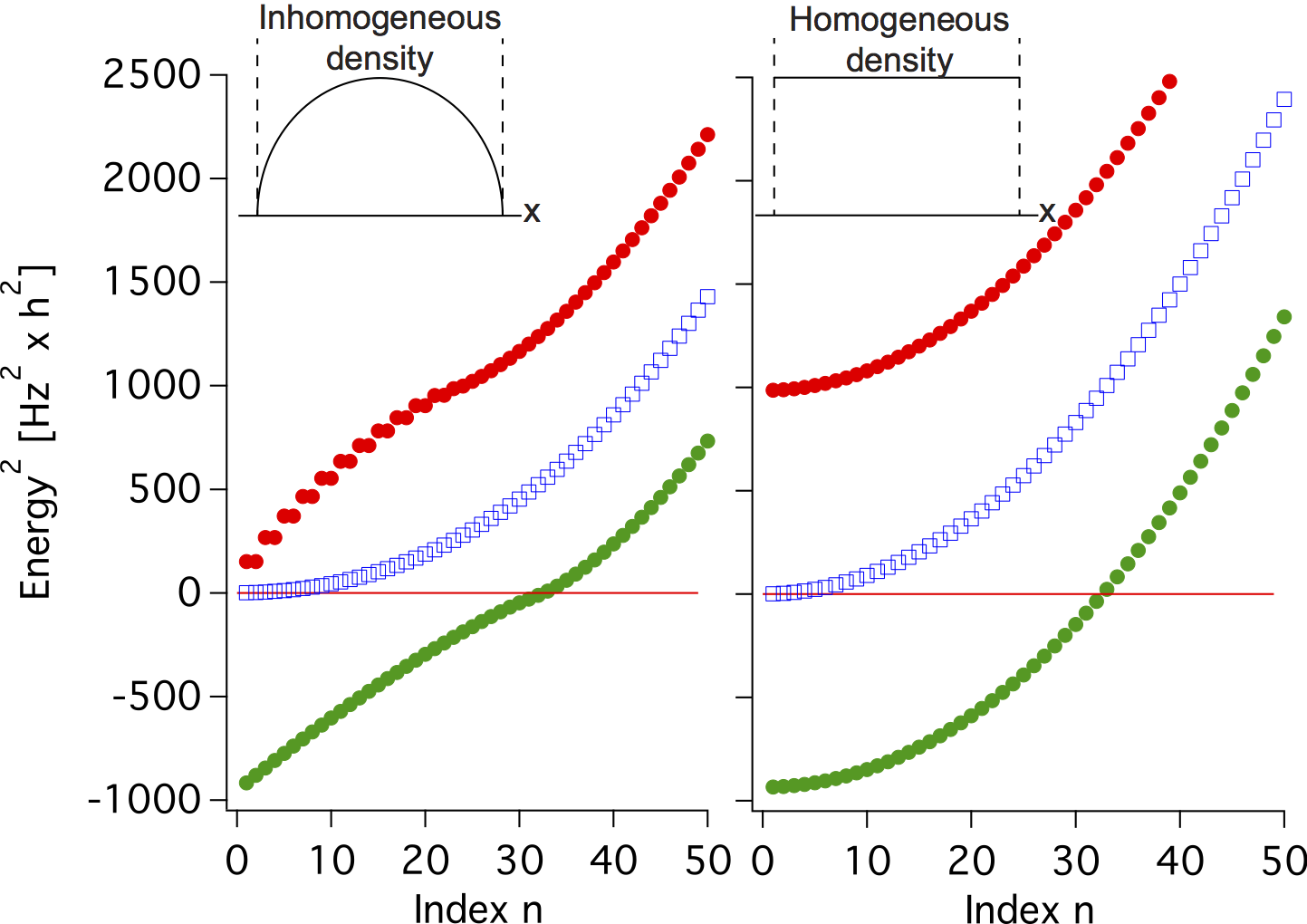}

\caption{(Color Online).  Excitation energies of spin modes.  Shown is the square of the numerically obtained eigenvalues $E_n; n = 1,2,3,...$ for inhomogeneous density profile (left panel) and homogeneous density (right panel).  Eigenvalues are plotted for q = + 5 Hz (red, filled circles), q = 0 (blue, open circles) and q = - 5 Hz (green, filled circles).  $E^2 < 0$ indicates an imaginary eigenvalue indicative of an instability.  Notice that the number of imaginary eigenvalues increases as $q$ becomes smaller.}
\label{fig:eigenvalues}
\end{figure}

We first discuss the numerical solutions for the eigenvalues before proceeding to the eigenvectors.  The left panel of Figure \ref{fig:eigenvalues} shows the eigenvalue spectrum computed for $p = 0$, for $q = +5, 0$ and $-5$ Hz, corresponding to stable, critically stable and unstable regimes.  The eigenvalue spectrum in the case of a homogeneous density gas at the same peak value of $n_0$ is plotted on the right panel for comparison.  In both cases we have plotted the square of the energy eigenvalue, $E^2$, rather the eigenvalue $E$ itself, since according to the Bogoliubov equations this quantity is always real for $p=0$.  The data in Figure \ref{fig:eigenvalues} highlight important differences between the two cases, as well as some of their similarities.

For $q>0$ (stable regime, $E^2>0$), the graph shows that the lowest eigenvalue (the mode with index $n=1$) is finite for both inhomogeneous and homogeneous densities, and thus the system always has an energy gap.  For a homogeneous system this gap is given by
\[ E_1 =  \sqrt{(\epsilon+q)(\epsilon+q+2 U)} \]
where $\epsilon$ is the lowest box mode.  For the parameters used, $E_1 = h \times 31$Hz.  For the inhomogeneous case the numerical data show that the energy gap is reduced by a factor of nearly 3, to $h \times 12$Hz.  As we will see later in Section \ref{sec:stable}, the lowest energy Bogoliubov eigenfunctions have a very different spatial profile for the two cases--for the inhomogeneous case, these modes are sharply localized near the Thomas-Fermi boundary (see Figure \ref{fig:stable_modes}), whereas for the homogeneous case they are mostly near the center of the cloud (see Eqn.\ (\ref{eq:mode})), which raises the homogeneous energies due to repulsive interactions with the $m = 0$ atoms.  With modes localized near the boundary rather than at the center, the energy difference between modes of opposite parity becomes negligible since they both have nearly the same overlap with the density profile $U(x)$.  Thus, the low-lying modes all come in nearly degenerate pairs that we term a ``parity doublet'' (the odd parity modes remain slightly higher in energy than the even parity ones, but the difference is very small for $n< \simeq 20$).  These doublets can be seen in the data as a series of pairs of dots.  For the homogeneous case, by contrast, the near degeneracy between even and odd states disappears, so there is no parity doublet.

For low $n$ the dispersion relation--the variation of $E$ with $n$--is also considerably different in the two cases.  For example, an inflection point can be seen at $n \simeq 25$ in the dispersion relation for the spin modes for the inhomogeneous distribution.  For $n<25$ the dispersion relation has negative curvature, similar to surface modes of a single component BEC \cite{pita03book}.  This stands to reason, as the modes are localized near the Thomas-Fermi surface.  For $n>25$ the curvature is positive, suggesting a transition to nearly free-particle behavior, although the eigenvalues are still smaller than the height of the potential $U_0$.  What determines the crossover point is still unclear, although the energy eigenvalue at the inflection point was observed empirically to increase with $q$.  

As $q$ becomes smaller, the entire spectrum of $E^2$ shifts to smaller values, until the lowest eigenvalue reaches zero at a point very close (within $\epsilon$) of $q=0$, which defines the boundary of the unstable region.  However, as we have plotted $E^2$ rather than $E$, the transition from stable to unstable regimes becomes a more smooth and continuous one, with the number of imaginary eigenvalues ($E^2 < 0$) increasing as $q$ decreases.  For example, at $q = - 5$ Hz 32 eigenvalues have become imaginary, while for $n \ge 33$ they still remain real.  Due to the tiny value of $\epsilon$, it plays no role, and the phase transition occurs at the same point for both inhomogeneous and inhomogeneous cases.  

\section{Comparison of theory with experiment} \label{sec:unstable}

We can separate the temporal dynamics into two phases.  In the early, growth phase, the population fraction in $\pm 1$, $f_{\pm 1}$, increases with time, but always remains small compared to that of the $m = 0$ state, $f_0$.  Bogoliubov theory can be used to study this phase.  The second, dynamical phase, occurs when all 3 components, $f_{+1},f_{-1},f_0$, have the same order of magnitude, and interact strongly with one another.  The dynamical phase is not captured by the Bogoliubov theory, but can be observed experimentally and compared with numerical simulations.  We discuss these two phases of the dynamics separately below.

\subsubsection{Growth phase}

\begin{figure} [htbp]
%\begin{array}
\includegraphics[width = 0.8\columnwidth]{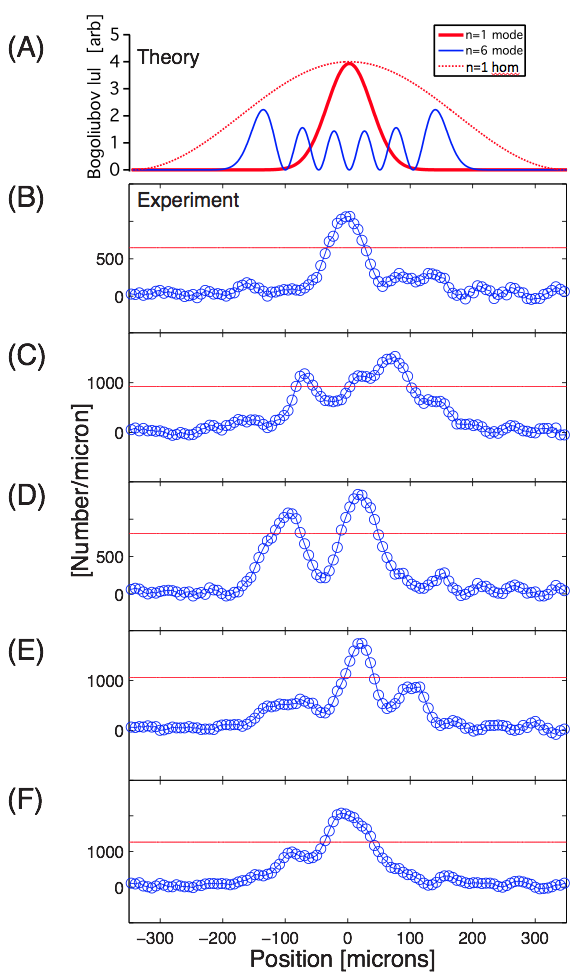} 
\includegraphics[width = \columnwidth]{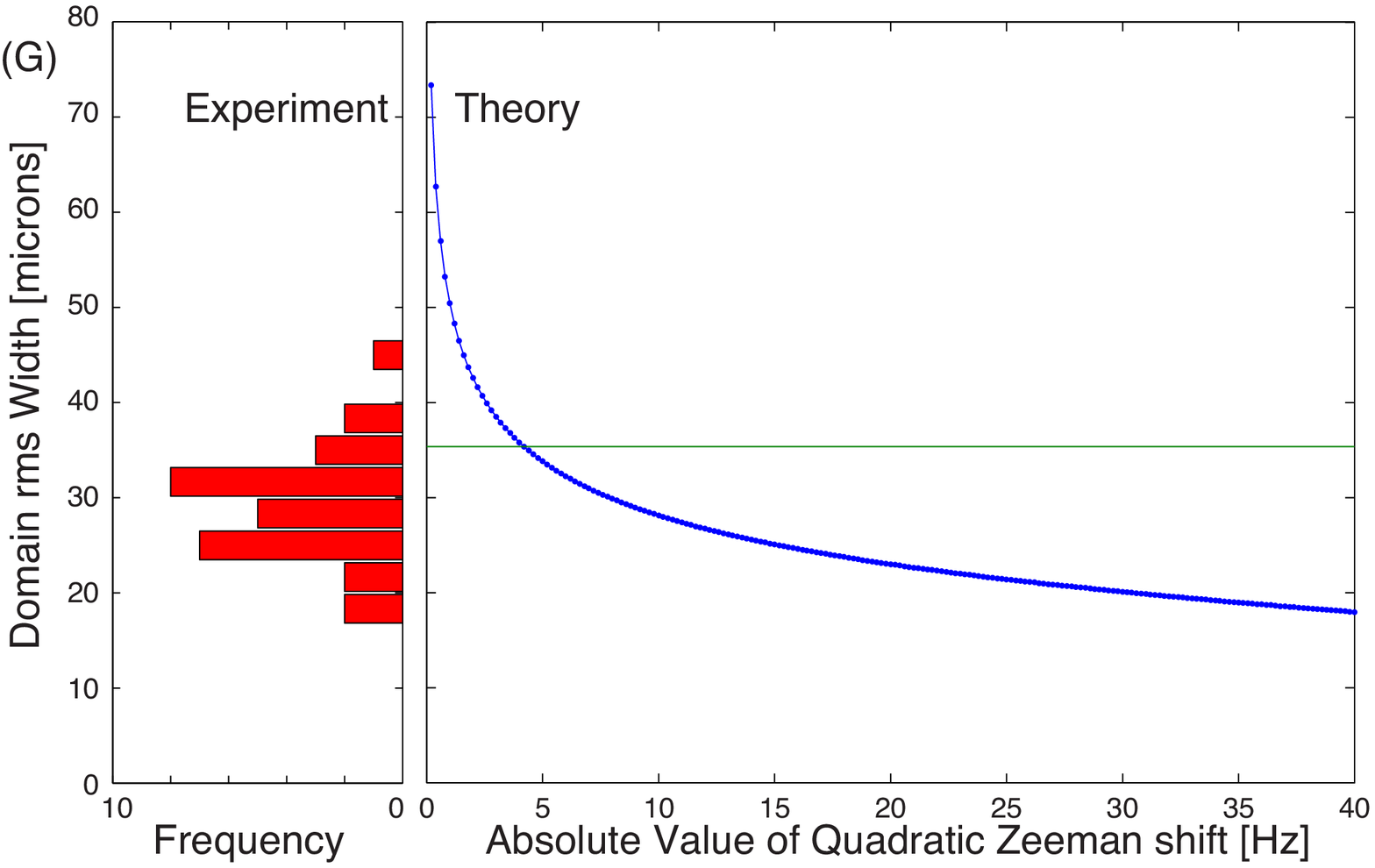} 
%\end{array}
\caption{(Color Online).  Unstable modes and their spatial profiles.  (A) Bogoliubov solutions for $U = 96$ Hz and $q = -4.2$ Hz.  Shown are the probability distributions of the maximally unstable mode ($n=1$, no nodes), a less unstable mode ($n=6$, 5 nodes), and the $n=1$ mode for the homogeneous density case.  (B-F) Representative experimental observations of domains nucleated by the instability at $t = 20$ ms after the quench.  Solid lines are the threshold used to determine rms domain sizes.  (G) (right) Calculated rms width of the maximally unstable eigenvector versus distance to the phase transition point $-q$.  The horizontal line is the value that correspnds to the experiment.  (left) Histogram of the observed domains for 30 runs of the experiment at $q=-4.2$ Hz shows an average domain size that is smaller than that of the lowest mode, suggesting the involvement of higher lying modes $n>1$.
\vspace{1in}}
\label{fig:unstable_modes1}
\end{figure}

In the growth phase, we computed the complex eigenvalues $E_n$.  These were sorted by the magnitude of their imaginary component, with $n=1$ mode having the largest imaginary component, $n=2$ the second largest, and so on.  Figure \ref{fig:unstable_modes1}A shows the numerically obtained solutions for the maximally unstable eigenvector, $n=1$, as well as that of a less unstable eigenvector, $n=6$, at a quadratic shift of $q = -4.2$ Hz, corresponding to the value used in the experiment.  The boundaries of the plot are the Thomas-Fermi surface, $x = \pm R_{TF} = \pm 340 \mu$m.  The maximally unstable mode is  localized near the center of the cloud, with a wavefunction resembling a Gaussian profile.  This is to be contrasted with Eqn.\ \ref{eq:mode}.  For larger $n$ the number of nodes increased, as did the spatial domain over which the mode function was non-zero.  

We define an rms width for each unstable mode as 
\begin{equation} 
x_n^{rms} = \sqrt{\frac{\int |u_n(x)|^2 x^2 dx}{\int |u_n(x)|^2 dx} }
\label{eq:rms}
\end{equation}
where $u_n$ is the $n$-th mode, and the upper and lower limits of integration are the Thomas-Fermi surface, $\pm R_{TF}$, respectively.  For the $n = 1$ mode, $x^{rms} = 0.05 R_{TF}$, nearly four times smaller than the homogeneous case, Eqn.\ \ref{eq:mode}, whose rms width is $0.18 R_{TF}$.  Thus the inhomogeneous density profile had a profound effect on the instability, causing the nucleation of localized spin domains.  

This tendency can also be envisioned by applying a local density approximation to the rate Eqn.\ (\ref{eq:Ehom}), which reflects the inhomogeneous gain profile for the spin-exchange process.  Since the $m = 0$ state is dynamically unstable for $q<0$, excitations develop at a rate that depends upon the {\em local} value of its density, $c_2 n_0(x)$.  Thus the spin domains become localized near the cloud center, where this rate is highest.  A similar effect was observed in reference \cite{Kang2017}, as noted earlier.

Another, equivalent way to view this is through consideration of how the instability amplifies spin noise, which can be represented in any basis.  Using plane waves with momentum ${\bf p}$, for example, and if the system contained noise that was uniformly distributed at all spatial frequencies up to $p_{max}/\hbar = 2 \pi / \xi_{sp}$, where $\xi_{sp} = 1.5 \mu$m is the spin healing length, it would begin in a delocalized state where the average amplitude was roughly the same everywhere.  This picture is consistent with the Truncated Wigner Approximation (TWA), as we discuss later.  As time develops, only the particular superposition of momentum eigenstates that reproduces the state $u_1(x) \equiv u_{MAX}(x)$ shown in Figure \ref{fig:unstable_modes1}A would be amplified significantly, and the spin distribution would develop into something that is spatially localized.  For longer times where the $m = 0$ component becomes depleted, this spatially localized state is no longer stable, but expands into a multi-domain structure.

Indeed, we can see from our experimental data in Figure \ref{fig:SG_slices} that the instability creates spin structures that are spatially localized near the center of the Thomas-Fermi region.   Through the second order correlation function we determined the average mode size to be $33 \mu$m, which is in good agreement with the rms width of $35 \mu$m predicted for the maximally unstable mode $u_{MAX}(x)$.  The latter is shown as a horizontal line in Figure \ref{fig:unstable_modes1}{G}, right panel.  

From Bogoliubov theory at $q = -4.2$ Hz, we estimate that 30 modes have an imaginary component.  Therefore, the maximally unstable mode is not the only active mode in the problem.  While the average domain size, as measured by the normalized second-order correlation function $g_2(x)$, is well captured by theory, we still must understand the varying number and size of the domains measured in the experiment.  To uncover these multi-mode effects, we look at the spread in experimentally observed domain sizes using an rms domain width analysis.  Figure \ref{fig:unstable_modes1}, panels (B-G), show mode profiles measured at the onset of the instability, $t = 20$ ms, when the mean population in the $\pm 1$ states was 15\%.  We show 5 separate  instances of the experimental quench sequence that are representative of the variations observed.  Shot to shot fluctuations reflected the stochastic dynamics discussed earler.  By observing peaks in the data, we could determine the rms size of domains associated with those peaks.  We determined these sizes using a simple criterion--the size was the minimum distance from the peak where the data crossed a threshold value of $ = e^{-1/2}$ of its peak value, as would be expected for the rms width of a Gaussian function.  The threshold is shown as a solid line in the plots.  The preponderance of multi-domain structures makes it clear that multiple modes are present.
%We did so by measuring the spatial coordinates at which the data, which measures the particle density $\propto |u|^2$, crossed a threshold value $ = e^{-1/2}$ of the peak value, as would be expected for a Gaussian function $u(x) = u_0 e^{-\frac{x^2}{4 (x^{rms})^2}}$.  The threshold is shown as a solid line in the plots.  The preponderance of multi-domain structures makes it clear that multiple modes are present.

Figure \ref{fig:unstable_modes1}G explores this multi-mode character of the instability by comparing the domain widths quantitatively.  On the right panel we have computed the rms width of the maximally unstable mode obtained from the Bogoliubov calculation, $u_{MAX}(x)$, versus the final quadratic Zeeman shift $|q| = -q$.  On the left panel we show a histogram of the observed rms sizes of the domains for 30 runs of the experiment at $q=-4.2$ Hz.  On the higher end, the distribution cuts off at a domain size that corresponds well with $x_{rms} = 35 \mu$m for the lowest mode, $u = u_{MAX}$.  Nonetheless, domains as small as $15 \mu$m were observed, not much larger than our spatial resolution of $10 \mu$m, indicating that the quench excited many higher order modes.  The tail in the distribution above $35 \mu$m is most likely caused by mistaken identification of multiple overlapping domains as a single, larger domain, an example of which is shown in panel (C).  

%We can separate the temporal dynamics into two phases.  First, there is a growth phase, during which the population fraction in $\pm 1$, $f_{\pm 1}$, increases with time, but always remains small compared to that of the $m = 0$ state, $f_0$.  Bogoliubov theory can be used to study the growth phase dynamics.  The second, dynamical phase, occurs when all 3 components, $f_{+1},f_{-1},f_0$, have the same order of magnitude, and interact strongly with one another.  The dynamical phase is not captured by the Bogoliubov theory, but can be observed experimentally.  

\begin{figure}[htbp]
\includegraphics[width=0.9\columnwidth]{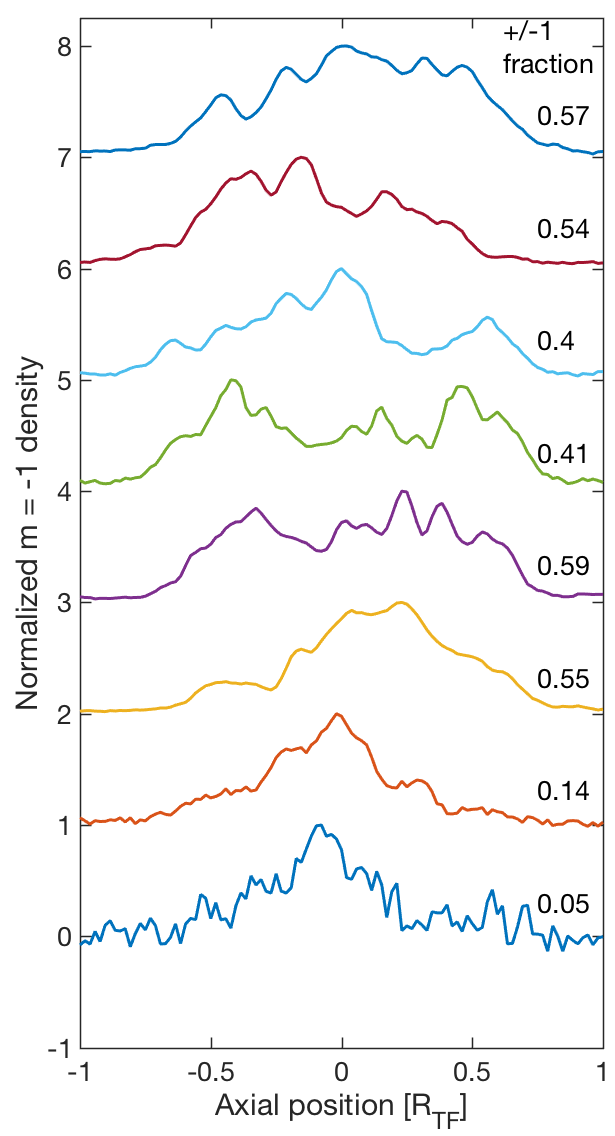}
\includegraphics[width=\columnwidth]{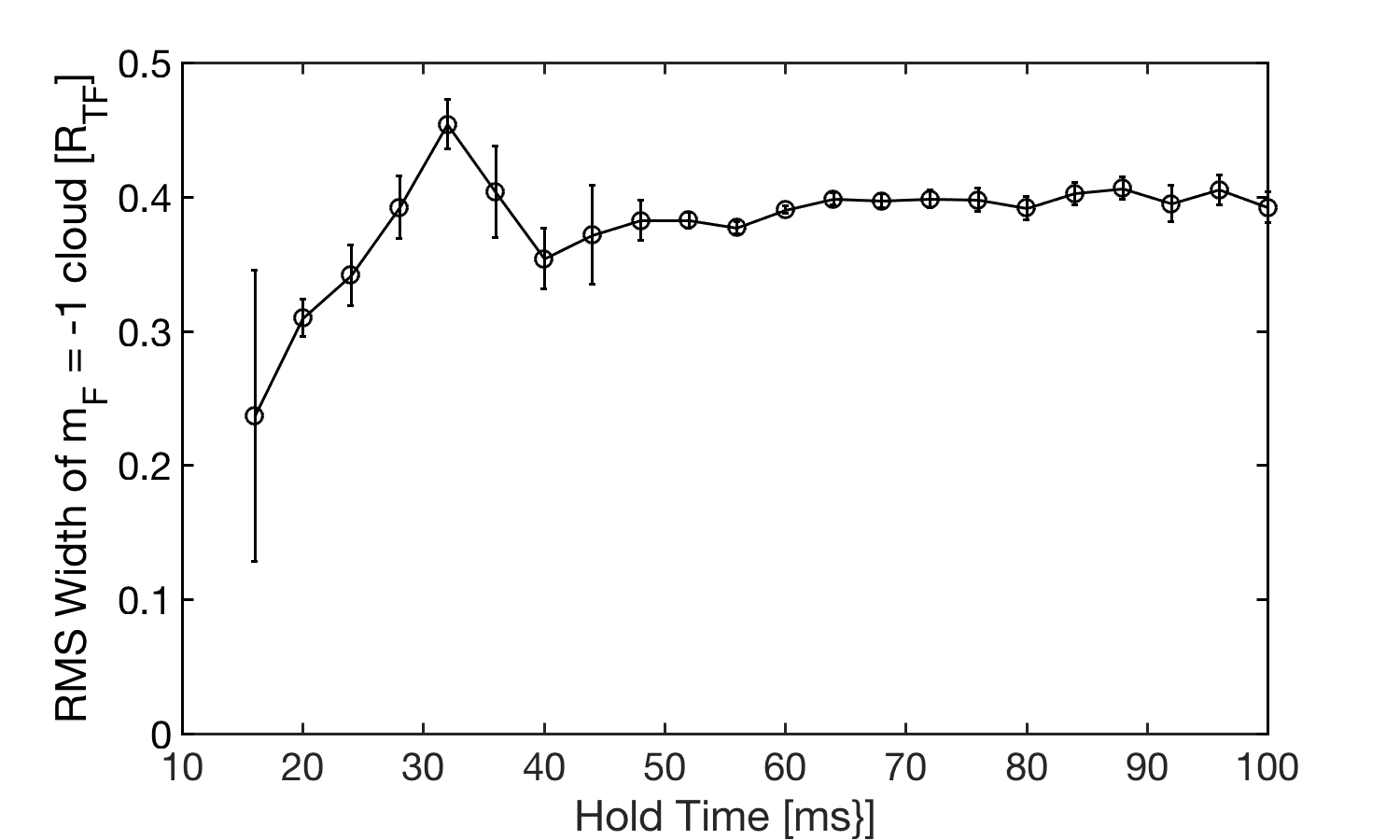}
\caption{(Color Online).  Evolving from growth to dynamical phases.  (Above) From bottom to top, experimental traces for the $m = -1$ state averaged over 3 shots to coarse grain over domain stochasticity, for times $t= 16$ to 44 ms after the quench in 4 ms intervals.  Corresponding fraction of atoms in the $\pm 1$ is written next to each curve.  Each curve has been normalized to its peak value and displaced for clarity of presentation.  (Below) RMS width of the density profile versus hold time.  Error bars are the standard deviation of the 3 separate measurements.}
\label{fig:averages}
\end{figure}

\subsubsection{Dynamical phase}

In this later phase, we observe a coarsening of the domains generated in the growth phase.  The upper panel of Figure \ref{fig:averages} shows the crossover from growth to dynamical phases in the one-dimensional $m = -1$ density profiles, which have been plotted against their axial position normalized to the axial Thomas-Fermi radius, $R_{TF}$.  Similar data were obtained for the $m = +1$ profile.  Each curve is an average of 3 experimental runs normalized to the peak value at each time step, with each curve displaced by 1 for clarity.  The average reduces the effect of stochastic fluctuations associated with spontaneous domain formation, allowing us to focus on the coarsening trend--there is a growth in the overall size of the $m = \pm 1$ clouds with time.  As seen in the figure, for short times, when Bogoliubov theory is still applicable, the density profile grows from the center of the cloud, forming a localized hump at a time when $f_{pm} \approx 0.05$.  As time increases, this hump grows in size to envelop the cloud.  

Although we can no longer use the unstable Bogoliubov eigenmodes to analyze the dynamical phase, we can  compare our data with numerical simulations.  To this end we compute a different rms width, this one pertaining to the entire $m = -1$ cloud,
\begin{equation} 
x^{rms} = \sqrt{\frac{\int |u(x)|^2 x^2 dx}{\int |u_n(x)|^2 dx} }
%\label{eq:rms}
\end{equation}
which is shown in the lower panel of Figure \ref{fig:averages}.  Here $|u(x)|^2$ is equal to the measured density profiles shown in the upper panel of the figure.  The width of the density hump is seen to increase with time as the system evolves into the dynamical phase, exhibiting an overdamped oscillation before reaching a steady-state value of about $0.4 R_{TF}$.  At $t = 16$ ms the super-Poissonian noise was larger than at later times, and imperfect averaging led to a larger error bar.  
%[Note:  the Thomas-Fermi density profile $n(x) = n_0(1-x^2/R_{TF}^2)$ has an rms value = $0.5 R_{TF}$, so the steady-state corresponds to 70\% of the cloud size].  
Although the coarse cloud size in each of the three interpenetrating quantum fluids, $m = 0$ and $m = \pm 1$, has reached a steady-state, the dynamics have not yet halted, as the fraction $f_\pm$ continues to steadily increase, as seen earlier in Figure \ref{fig:diagram}C.  In this later phase of the non-equilibrium behavior micro-domains still exist and move throughout the cloud.  As noted in our earlier work, one can observe a small, local magnetization density $M(x) = n_{+1}(x)-n_{-1}(x)$.  Thus the $m = \pm 1$ clouds eventually separate from one another \cite{vinit2013}.

\begin{figure*}[htbp]
\begin{center}$
\begin{array}{ccc}
\includegraphics[width= 0.75\columnwidth]{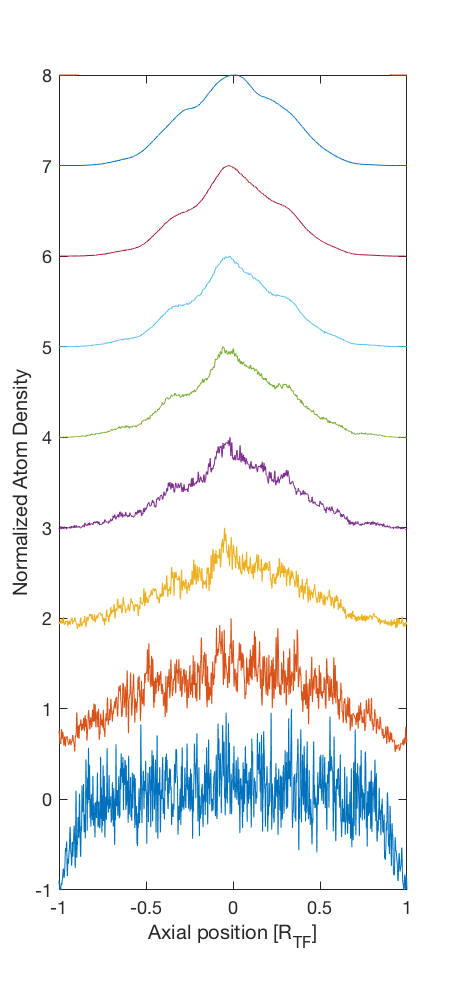}&
\hspace{-0.25in}
\includegraphics[width= 0.75\columnwidth]{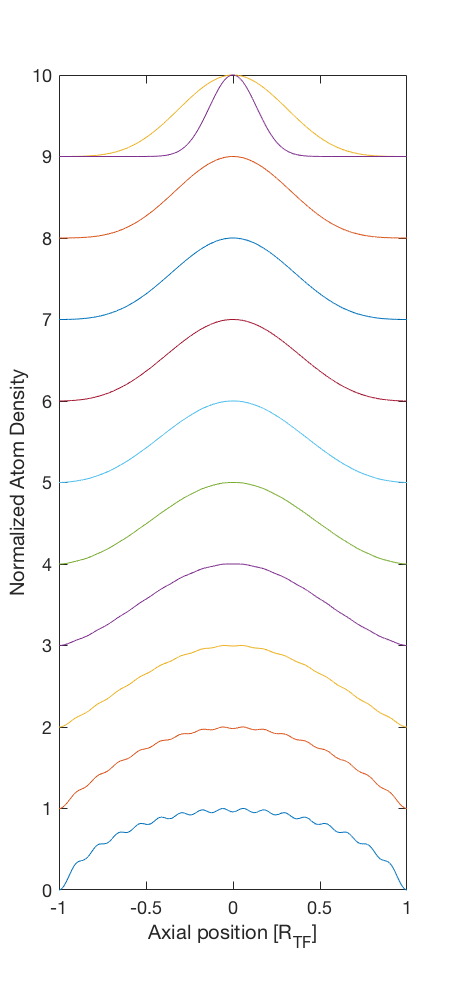}
\end{array}$
\end{center}
\caption{(Color Online).  Average density profiles predicted by theory.  Numerical simulations are shown on the left.  Bogoliubov theory, Eqn.\ (\ref{eq:expansion}) is shown on the right, for times $t/\tau_{MAX} = 1,2,...10$, with the maximally unstable eigenvector, $|u_{MAX}|^2$, shown in the uppermost graph for comparison.  Here $\tau_{MAX}= 1/{\rm Im}(E_{MAX})$ is the time scale associated with $u_{MAX}$.  In both panes, each curve has been normalized to its peak value and displaced for clarity of presentation.}
\label{fig:averages2}
\end{figure*}

\subsubsection{Numerical results}

Numerical simulations allowed us to bridge the growth and the dynamical phases of evolution, and to probe the early part of the growth phase where little experimental data could be obtained.  We performed one-dimensional simulations of the 3 coupled spinor Gross-Pitaevskii (GP) equations seeded with noise according to the Truncated Wigner Approximation, or TWA.  The population dynamics derived from the simulations were reported in our earlier work \cite{vinit2013}.  Here we provide more details including the numerically obtained wavefunctions and their temporal evolution.  Our numerical procedure is a straightforward forward time propagation of the equations of motion using a time-splitting spectral method (see, for example, reference \cite{Bao2003} and references therein).  The TWA approximation is expected to be valid for both short and long times, as long as the initial condition is a classical state \cite{blakie2008,polkovnikov2010,sau2009,barnett2011}.  To implement these simulations, as discussed in \cite{vinit2013}, we assumed a BEC initially at zero temperature and obtained the initial wavefunction for the $m = 0$ component numerically.  Vacuum noise in the $m = \pm 1$ states was simulated as classical noise, and we computed the average density, $\langle \psi_m^\dag \psi_m \rangle$, as an ensemble average over 30 separate simulations using different random initial conditions.  Vacuum modes with wavelength less than $\xi_{spin}$ are not expected to contribute to the spin instability.  Therefore, we imposed a cutoff energy of $c_2 n_0$, which resulted in $N_v \simeq $700 virtual particles, while the condensate contained $5 \times 10^6$ particles, similar to the experimental conditions.  To study the early time behavior in the simulations it was also essential to subtract a constant from the average density $ = N_v/(2 R_{TF})$ equivalent to the sum of all virtual particles added, which was done according to the Weyl representations of the field operators \cite{polkovnikov2010}.   

As mentioned earlier, the experimental data measure the integrated column density $\tilde{n}(x) = \int n(x,y,z) dy dz $.  To compare this with a one-dimensional simulated density profile $n(x)$, we first posit a solution that is separable in space between axial ($x$) and transverse ($y,z$) coordinates:
\begin{equation}
\Psi_m(x,y,z,t) = \psi_m(x,t) \xi(y,z) 
\label{eq:separable}
\end{equation}
where $\psi_m(x,t)$ is the simulated wavefunction for spin state $m$.  Since the quench is one-dimensional, we may assume that the transverse mode function $\xi$, normalized as $\int |\xi|^2 dy dz = 1$, is time-independent as the dynamics are frozen.  With the approximation Eqn.\ (\ref{eq:separable}) and the three-dimensional density distribution $n_m(x,y,z) = |\Psi_m|^2$, the measured column density becomes
\[ \tilde{n}_m(x) = \int n_m(x,y,z) dy dz = |\psi_m(x,t)|^2 \]
and is identical to the one-dimensional density profile obtained from the simulation.  However, for a Thomas-Fermi BEC, the solution is not strictly separable, as the transverse Thomas-Fermi radius depends on axial position $x$.  In effect, Eqn.\ (\ref{eq:separable}) assumes a Thomas-Fermi cylinder rather than a cigar.  If the cigar aspect ratio is very large (70 in our case), the difference between cylinder and cigar is insignificant, particularly since most of the important quench dynamics occur near the cloud center.  However, care must be paid when one approaches the axial Thomas-Fermi radius, $x = \pm R_{TF}$, as there are likely to be quantitative differences.  

On the left panel of Figure  \ref{fig:averages2}, we show the result of the numerical simulations, and on the right pane is shown the prediction for the ensemble averaged density profile from Bogoliubov theory.  Here we have computed $\sum |u_k|^2 e^{\Gamma_k t}$ where the sum runs over all unstable modes.  The growth factor $\Gamma_k = {\rm Im} [E_k/h]$.  The curves have been normalized in the same manner as for the experiment.  At $t=0$ the exponential factor is 1 for all modes, resulting in a nearly uniform initial density profile.  

Both Bogoliubov and full numerical theories agree with one another during the growth phase of the dynamics.  Moreover, the theory confirms the local density picture discussed earlier, where a uniform distribution comprised of vacuum fluctuations in all modes becomes narrower with time, eventually becoming localized near the cloud center.  For the Bogoliubov results, at later times the maximally unstable mode, $\Gamma_{MAX}$, begins to dominate, and the curves begin to peak around this mode function, $u_{MAX}$, which is shown as the narrow distribution in the uppermost plot of the right pane.  Even at a time $t = 10/\Gamma_{MAX}$, however, the Bogoliubov density distribution has not fully converged to $u_{MAX}$, but remains broader.  Our experimental data shows the formation of the localized structure, but not its precursor, the uniform phase, which is hidden in experimental noise.  The uniform phase is, however, captured by the TWA simulations (see the lowest traces of the left panel).  

An interesting artifact in the simulations can also be observed in Figure \ref{fig:averages2}, one which illustrates some of the limitations of the Bogoliubov analysis.  The TWA initial condition was taken to be a sum of Bogoliubov eigenmodes prior to the quench (see the next section for details of these stable modes), with random coefficients.  Since these modes are defined on $x \in [-R_{TF},+R_{TF}]$, their amplitude goes exactly to zero at the Thomas-Fermi radius.  However, the numerical simulations are not restricted to the Thomas-Fermi volume, but capture the full details of the cloud's surface structure, even for $|x| >  R_{TF}$.  Thus at short times in the simulation, the repulsive interaction between atoms redistributed the $m = \pm 1$ density from inside to outside of $R_{TF}$ such that as $x$ approached $\pm R_{TF}$ from within the cloud, the Weyl correction was no longer accurate and yielded a negative density, seen in the lowest traces.  Therefore, at short times the $m = \pm 1$ density should be even more uniform than the simulation suggests.  This artifact had no bearing on the simulation at longer times, since the Weyl correction, which counts only the vacuum fluctuations, was insignificant in comparison with the number of real particles.  Nonetheless, it illustrates the difficulty in describing the details of the dynamics near the Thomas-Fermi surface.  
\begin{figure} [htbp]
\includegraphics[width = \columnwidth]{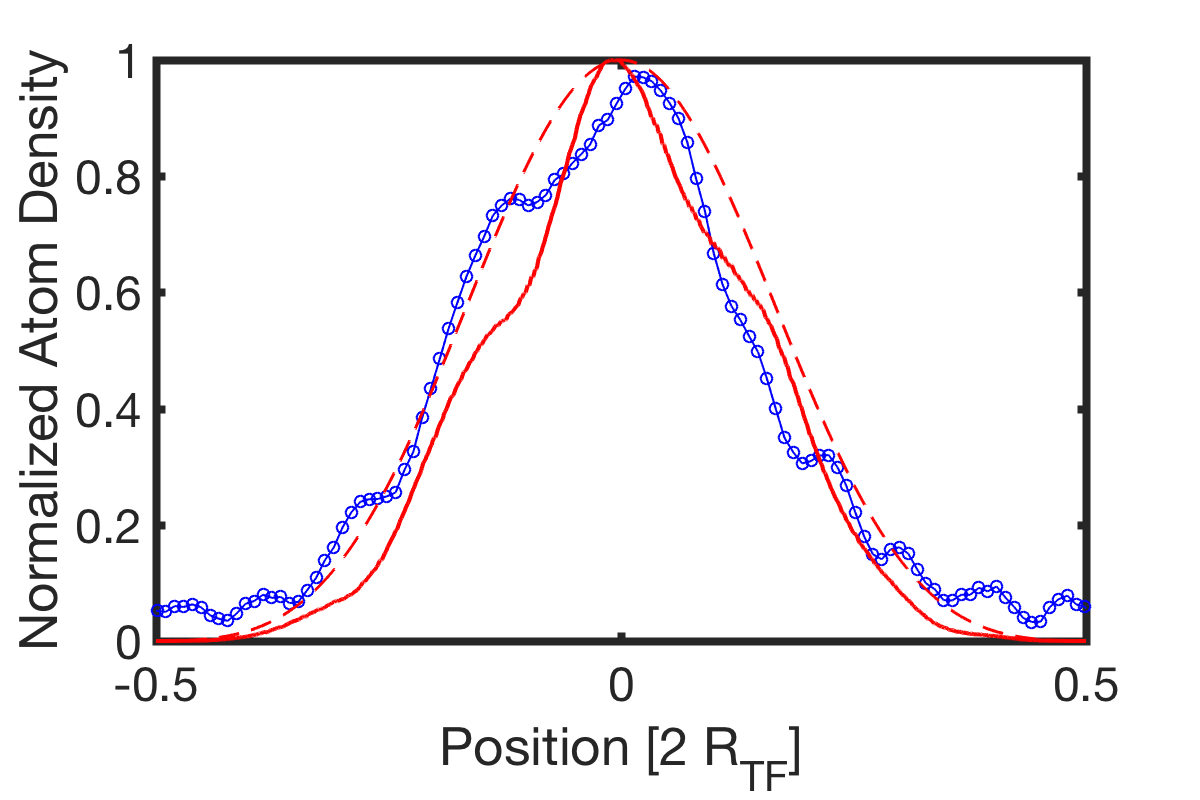}
\caption{(Color Online).  Average density profiles for experiment (open circles), simulation (thick red line), and Bogoliubov theory (dashed line).}
\label{fig:averages3}
\end{figure}

Finally, an agreement between all 3 density profiles--the experimental data, TWA simulations, and Bogoliubov theory--was found at a time approaching the crossover from growth phase to dynamical phase.  This data is shown in Figure \ref{fig:averages3}.   The experiment and TWA curves were taken at a time when the $\pm 1$ fractions were similar, having reached $f_{\pm1}=0.15$ and $0.17$, respectively.  This allowed us to circumvent a factor of 4 difference observed in the absolute timescale for the quench dynamics between the two, as noted in \cite{vinit2013}.  These data contain residual oscillations due to imperfect averaging.  The Bogoliubov theory was taken at a time $t = 10 \tau_{MAX}$.  All 3 curves are broader than the maximally unstable eigenmode $u_{MAX}$.  
%All 3 curves are narrower than the Thomas-Fermi density profile, which is shown for comparison, but still broader than the maximally unstable eigenmode $u_{MAX}$.  
Bogoliubov theory should converge to the maximally unstable eigenmode; however, this only occurs after a sufficiently long time.  We observed convergence at a time $t \approx 100 \Gamma_{MAX}$, by which time in the experiment the $m = 0$ cloud would have been significantly depleted, violating the Bogoliubov approximation.  This provides further confirmation that our experiment is in a multi-mode regime even during its growth phase, when Bogoliubov theory is valid.  

\begin{figure} [htbp]
\includegraphics[width=\columnwidth]{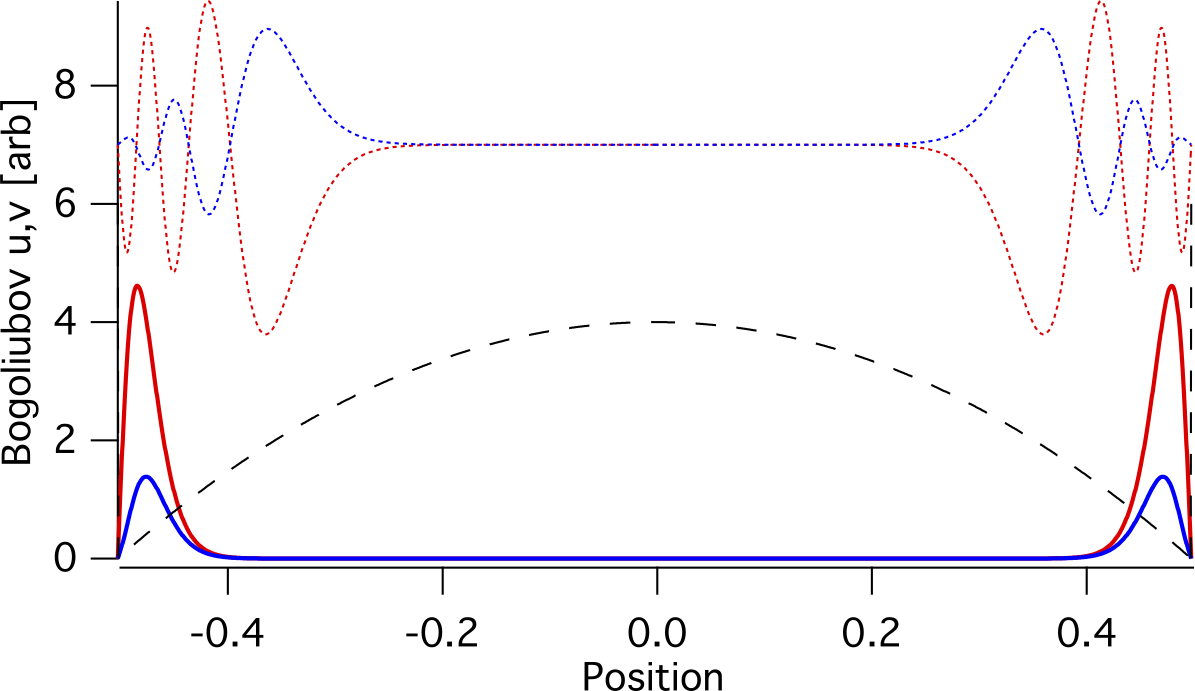}
\caption{(Color Online).  Stable spin modes.  Shown are Bogoliubov functions $u(x)$ (red) and $v(x)$ (blue) for $q = + 5$ Hz, $U = 96$ Hz, for the inhomogeneous Thomas-Fermi density profile, whose repulsive potential is shown as a black dashed line.  (Below, solid lines) Mode with the lowest energy $n = 1$.  (Above, dotted lines) A higher excitation mode $n = 9$, offset vertically for clarity.}
\label{fig:stable_modes}
\end{figure}

\subsection{Stable Modes} \label{sec:stable}

For completeness, we include a discussion of the collective excitation spectrum for $q > 0$, although no experimental data was taken in this regime.  Nonetheless, it provides additional insights into the difference between homogeneous and inhomogeneous cases, particularly as $q \rightarrow 0$, the phase transition point.  Here the eigenvalues are all real and positive, and $n$ is a mode index by which they are sorted in increasing order.  Similar to the unstable modes discussed in the previous section these modes also depart from the homogeneous Bogoliubov solutions, Eqn.\ (\ref{eq:box}).  Their spatial profile also depends upon $q$ in a manner that was determined numerically.  

The lower graph of Figure \ref{fig:stable_modes} shows the numerically obtained mode functions, $u_1(x)$ and $v_1(x)$, for the lowest energy eigenvalue $E_1$, at a quadratic Zeeman shift of $q = +5$ Hz.  These functions are sharply localized near the Thomas-Fermi boundary at $x/L_{TF} = \pm 1/2$, in stark contrast with the homogeneous modes that are delocalized throughout the Thomas-Fermi region.  For increasing $n$ the modes penetrate further into the cloud--for comparison, the $n = 9$ mode, with even parity, is shown in the upper graph.  

\begin{figure} [htbp]
\includegraphics[width= \columnwidth]{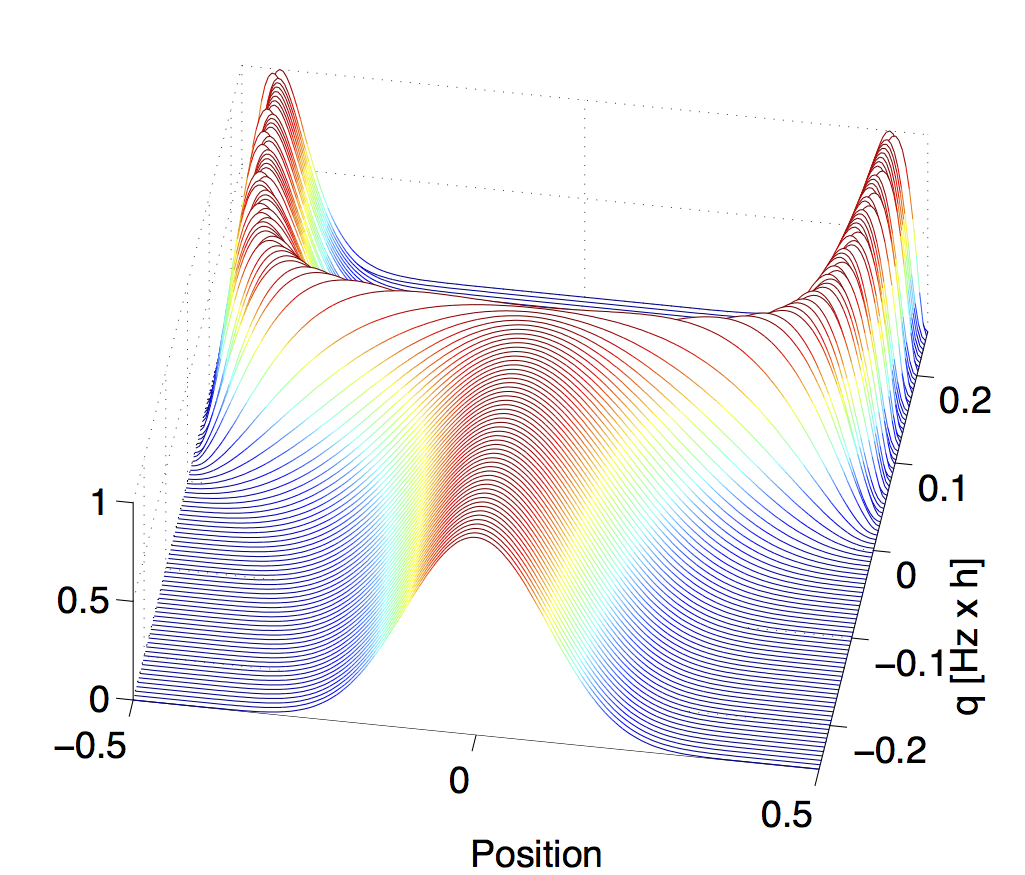}
\caption{(Color Online).  Transformation of the wavefunction caused by the instability.  Shown is the numerically obtained function $|u(x)|^2$ for the lowest energy eigenvalue for 100 values of $q$ near the phase transition point, $q_{crit} =-0.005$ Hz.  The deviation of $q_{crit}$ from 0 occurs due to the discreteness of the eigenvalues.  The data are scaled to the peak value of $|u(x)|^2$ at each $q$ for clarity of presentation.  For positive $q$ the excitation is localized at the Thomas-Fermi boundary of the $m = 0$ cloud, while for negative $q$ it is localized in the cloud center.  The transition from the boundary to the center of the cloud occurs very suddenly as $q$ is changed--note that the entire span of $q$ is only 0.4 Hz in the figure.}
\label{fig:stable_unstable}
\end{figure}

We can understand the mode structure for $q>0$ in terms of the total potential appearing in the Bogoliubov equations:
\[ U(x) = c_2 n_0(x) = c_2 n_0  \left ( 1-\frac{x^2}{R^2} \right ) \]
where $n_0$ is the density averaged over the 2 radial directions of the optical trap.  Beyond the Thomas-Fermi radius we can set $U(x) = \infty$, since the harmonic potential increases very rapidly in comparison with the energy scale $c_2 n_0$.  Figure \ref{fig:stable_modes} shows $U(x)$ as a dashed line.  It resembles the box potential obtained in the homogeneous case, but with an additional bump at $x = 0$ that shifts the excitations away from the cloud center and toward the Thomas-Fermi boundary.  This repulsion was discussed earlier as being due to antiferromagnetism--for $c_2 > 0$ the spin $m = 0$ and spin $|m| = 1$ quantum fluids repel one another.  Since the excitations are $\pm 1$ atom pairs, their minimum energy configuration is a localized state at the edge of the cloud.

Crossing the phase transition results in a transformation from stable to unstable behavior of the eigenmode.  This effect is explored in Figure \ref{fig:stable_unstable} for the lowest energy state.  Only within a tiny region near the critical point having a width of $0.1$ Hz, does the mode function become delocalized.  By contrast, in the homogeneous case the mode  $\sin{\pi(x+L/2)/L}$  remains the same on both sides of the phase transition, and is always maximum at $x=0$.

\section{Conclusion}

For spatially extended quantum systems, the study of relaxation toward equilibrium naturally involves the dynamics of many modes and the flow of energy between them.  We have used the second-order correlation function, $g_2(x)$, and statistical analysis of domain widths to reveal the richness of this multi-mode behavior in quenched antiferromagnetic spinor Bose-Einstein condensates.  These approaches, combined with a portfolio of theoretical tools, have allowed us to span the data from the early, growth phase, to the later, dynamical phase.  For the former case, Bogoliubov theory is a semi-analytical approach which provides much physical intuition.  However, for the latter case we have relied on numerical simulations in order to explain the data.  Future work will explore possible dynamical universality in the long time dynamics, which could provide new analytical insights beyond the numerical work that has been done here \cite{Karl2017}.

This work was supported by NSF grant No.\ 1707654.

%\bibliographystyle{natbib}
%\bibliography{References,References_repaired}

\end{document}